\documentclass[nofootinbib, aps, prd, a4paper, 10pt, superscriptaddress, eqsecnum, showkeys]{revtex4-2}
\usepackage[a4paper, top=2cm, bottom=2cm, left=2.5cm, right=2.5cm]{geometry}

\usepackage{amsmath}
\usepackage{amsfonts}
\usepackage{amsthm}
\usepackage{bm}
\usepackage{mathrsfs}
\usepackage{braket}

\usepackage{graphicx}
\usepackage{booktabs}

\usepackage{xcolor}
\usepackage[pdfusetitle]{hyperref}
\hypersetup{colorlinks = true, allcolors = blue}

\usepackage{orcidlink}

\newcommand{\bea}{\begin{eqnarray}}
\newcommand{\eea}{  \end{eqnarray}}

\newcommand\be{\begin{equation}}
\newcommand\ee{\end{equation}}

\newcommand{\ba}{\begin{eqnarray}}
\newcommand{\ea}{\end{eqnarray}}
\allowdisplaybreaks[4]

\allowdisplaybreaks[4]

\begin{document}

\title{Adiabatic Perturbations in GW170817-Compatible Einstein-Gauss-Bonnet Inflation}
\author{S.D.~Odintsov\orcidlink{0000-0002-3529-7030}}
\email{odintsov@ieec.cat} \affiliation{ICREA, Passeig Luis
Companys, 23, 08010 Barcelona, Spain} \affiliation{Institute of
Space Sciences (ICE-CSIC) C. Can Magrans s/n, 08193 Barcelona,
Spain}
\author{V.K. Oikonomou\orcidlink{0000-0003-0125-4160}}
\email{voikonomou@gapps.auth.gr} \affiliation{Department of
Physics, Aristotle University of Thessaloniki, Thessaloniki 54124,
Greece} \affiliation{Center for Theoretical Physics, Khazar
University, 41 Mehseti Str., Baku, AZ-1096, Azerbaijan}

\begin{abstract}
We study the adiabaticity of the cosmological perturbations in the
context of inflationary Einstein-Gauss-Bonnet theories. We focus
on viable inflationary Einstein-Gauss-Bonnet theories which are
compatible with the current Cosmic Microwave Background radiation
experiments and also are compatible with the GW170817
observations. We derive the effects of the adiabaticity
requirement on the Einstein-Gauss-Bonnet physical parameters and
we show that the sound speed of the scalar perturbations and the
propagation speed of the tensor perturbations are constrained. We
consider two classes of inflationary viable and
GW170817-compatible theories, and in the first class the
adiabaticity is not violated during inflation, while in the second
class the adiabaticity is violated only at the end of inflation.
We discuss the effects of the adiabaticity violation in the second
class of models. From our analysis, it seems that only one class
of viable EGB inflationary theories, which is also compatible with
the GW170817 event, is free from adiabaticity pathologies.
\end{abstract}

\maketitle

\section{Introduction}

The next decade is expected to be quite fascinating, since major
experiments having to do with cosmology are expected to commence.
The most important aim of these experiments is to reveal whether
the inflationary regime of our Universe took place primordially.
Undoubtedly, the primordial era of our Universe has long been
theorized and the only theory that can actually be tested is
inflation
\cite{inflation1,inflation2,inflation3,inflation4,inflation5,inflation6}.
The inflation theory is a theoretical construction that solves the
flatness and horizon problems of the standard Big Bang cosmology
and currently is constrained by the Cosmic Microwave Background
(CMB) experiments. A smoking gun signature of inflation would be
the observation of the B-modes in the CMB polarization. This is
the aim of the near future CMB experiments like the Simons
Observatory \cite{SimonsObservatory:2019qwx} and the LiteBird
experiment \cite{LiteBIRD:2022cnt}. Apart from the CMB, the
inflationary era can be indirectly revealed by the existence of a
stochastic gravitational wave background with cosmological origin,
and this will be probed in the mid 2030 by the future
gravitational wave experiments
\cite{Hild:2010id,Baker:2019nia,Smith:2019wny,Crowder:2005nr,Smith:2016jqs,Seto:2001qf,Kawamura:2020pcg,Bull:2018lat,LISACosmologyWorkingGroup:2022jok}.
NANOGrav already confirmed the existence of a stochastic
gravitational wave background in 2023 \cite{NANOGrav:2023gor}, but
it is highly unlikely that such a background can be attributed
solely to inflation \cite{Vagnozzi:2023lwo,Oikonomou:2023qfz}.

Apart from the single scalar field modelling of inflation
\cite{inflation1,inflation2,inflation3,inflation4,inflation5,inflation6},
there is also the modified gravity description
\cite{reviews1,reviews2,reviews3,inflation6}, and two mainstream
modified gravity theories are of fundamental importance for
modelling inflation, $F(R)$ gravity
\cite{Nojiri:2003ft,Capozziello:2005ku,Capozziello:2004vh,Capozziello:2018ddp,Hwang:2001pu,Bamba:2014mua,Nojiri:2006gh,Song:2006ej,Capozziello:2008qc,Bean:2006up,Capozziello:2012ie,Faulkner:2006ub,Olmo:2006eh,Sawicki:2007tf,Faraoni:2007yn,Carloni:2007yv,
Nojiri:2007as,Capozziello:2007ms,Deruelle:2007pt,Appleby:2008tv,Dunsby:2010wg,Oikonomou:2020oex,Oikonomou:2020qah}
and Einstein-Gauss-Bonnet (EGB) gravity
\cite{Hwang:2005hb,Cognola:2006sp,Nojiri:2005vv,Nojiri:2005jg,Satoh:2007gn,Yi:2018gse,Guo:2009uk,Guo:2010jr,Jiang:2013gza,vandeBruck:2017voa,Pozdeeva:2020apf,Vernov:2021hxo,Pozdeeva:2021iwc,Fomin:2020hfh,DeLaurentis:2015fea,Chervon:2019sey,Nozari:2017rta,Odintsov:2018zhw,Kawai:1998ab,Yi:2018dhl,vandeBruck:2016xvt,Maeda:2011zn,Ai:2020peo,Easther:1996yd,Codello:2015mba,Oikonomou:2021kql,Oikonomou:2022xoq,Odintsov:2020sqy,Oikonomou:2024etl,Fier:2025huc,Pozdeeva:2024ihc,Odintsov:2026cxz}.
Both these theories are very well theoretically motivated, because
the EGB term and the $F(R)$ gravity emerge as the one-loop
corrections of a single scalar field in its vacuum configuration,
which is conformally or minimally coupled. Indeed, the quantum
corrections of the scalar field action has the form
\cite{Codello:2015mba},
\begin{align}\label{quantumaction}
&\mathcal{S}_{eff}=\int
\mathrm{d}^4x\sqrt{-g}\Big{(}\Lambda_1+\Lambda_2
\mathcal{R}+\Lambda_3\mathcal{R}^2+\Lambda_4 \mathcal{R}_{\mu
\nu}\mathcal{R}^{\mu \nu}+\Lambda_5 \mathcal{R}_{\mu \nu \alpha
\theta}\mathcal{R}^{\mu \nu \alpha \theta}+\Lambda_6 \square
\mathcal{R}\\ \notag &
+\Lambda_7\mathcal{R}\square\mathcal{R}+\Lambda_8 \mathcal{R}_{\mu
\nu}\square \mathcal{R}^{\mu
\nu}+\Lambda_9\mathcal{R}^3+\mathcal{O}(\partial^8)+...\Big{)}\, ,
\end{align}
with the parameters $\Lambda_i$, $i=1,2,...,6$ being dimensionful
constants. Doubts were cast on EGB theories and Horndeski theories
in general back in 2017, with the release of the GW170817
observation
\cite{TheLIGOScientific:2017qsa,Monitor:2017mdv,GBM:2017lvd,LIGOScientific:2019vic},
see Refs.
\cite{Ezquiaga:2017ekz,Baker:2017hug,Creminelli:2017sry,Sakstein:2017xjx,Boran:2017rdn}.
The GW170817 constraint on the gravitational wave speed is,
\begin{align}
\label{constraintGW170817} \left| c_T^2 - 1 \right| < 6 \times
10^{-15}\, ,
\end{align}
and since EGB and many Horndeski theories in general predict a
speed of tensor perturbations distinct from unity in natural
units, these theories were cast out from being viable inflationary
theories. A remedy however for EGB theories was proposed in Refs.
\cite{Oikonomou:2021kql,Oikonomou:2022xoq,Odintsov:2020sqy,Oikonomou:2024etl},
which solved the viability issues of EGB theories.

In this work we focus on the viable EGB inflationary theories of
Refs.
\cite{Oikonomou:2021kql,Oikonomou:2022xoq,Odintsov:2020sqy,Oikonomou:2024etl}
and we study an aspect of these theories that has never been
formally addressed in the literature, namely the adiabaticity of
the cosmological perturbations of these theories. Specifically, we
shall study two distinct classes of viable EGB theories, the
constrained EGB inflationary theories developed in
\cite{Oikonomou:2021kql,Oikonomou:2022xoq,Odintsov:2020sqy} and
the unconstrained EGB inflationary theories
\cite{Oikonomou:2024etl}. We shall analyze the adiabaticity
criteria for general inflationary EGB theories, and we shall study
which of the two classes respects the adiabaticity criteria. As we
demonstrate, the constrained EGB theories violate the adiabaticity
criteria at the end of the inflationary era, in the last few
$e$-foldings, thus particle creation takes place at the end of
inflation, but also the power spectrum is affected by the
non-adiabaticity of the perturbations. However, the unconstrained
EGB theories do not violate the adiabaticity criteria.

For the rest of this work, we assume that the spacetime is
described by a flat Friedmann-Robertson-Walker (FRW) spacetime,
with line element,
\begin{equation}
\label{JGRG14} ds^2 = - dt^2 + a(t)^2 \sum_{i=1,2,3}
\left(dx^i\right)^2\, ,
\end{equation}
where $a(t)$ is the scale factor and also the Hubble rate is
$H=\frac{\dot{a}}{a}$.

\section{Adiabatic Perturbations and the Adiabatic Vacuum of EGB Theories}

\subsection{Adiabatic Cosmological Perturbation Definition}

In this section we thoroughly discuss the meaning of the adiabatic
vacuum and the adiabatic perturbations in cosmology. The
terminology ''adiabatic'' in curved spacetime quantum field
theories refers essentially to the quantum mechanical adiabaticity
and not the thermodynamical adiabatic processes. The term
adiabatic means essentially that the transitions between
instantaneous eigenmodes is suppressed, and no Bogoliubov mixing
occurs. If no Bogoliubov mixing occurs in the perturbation modes
of a cosmological theory, no particle production occurs and vice
versa. This occurs in cosmological theories where the background
parameters vary very slowly compared to the oscillation time scale
of the field modes.

In cosmological theories, the perturbations satisfy an evolution
equation of the form,
\begin{equation}\label{mainequationperturb}
\frac{\partial^2 \chi_k(\eta)}{\partial
\eta^2}+\omega_k(\eta)^2\chi_k(\eta)=0\, ,
\end{equation}
where $\chi_k(\eta)$ is $\chi_k(\eta)=a(\eta)\phi_k(\eta)$ and
$\phi_k(\eta)$ are the Fourier modes of the perturbations,
$a(\eta)$ is the scale factor and $\eta$ is the conformal time.
More importantly, the frequency $\omega_k(\eta)$ is a
time-dependent variable which is model dependent and depends
strongly on the Mukhanov-Sasaki variable in cosmological contexts.

In curved spacetime, the definition of a particle and the very own
definition of the vacuum state becomes quite fuzzy. However, the
adiabatic vacuum bypasses these problems of the Fock space of the
perturbations, and it relies heavily on the high frequency regime
of the perturbations, the subhorizon regime. The main problem with
Eq. (\ref{mainequationperturb}) is that the frequency
$\omega_k(\eta)$ can get quite complicated and thus the definition
of an adiabatic vacuum can get cumbersome, and particles may be
produced if the solutions are not adiabatic, due to the mixing of
the in- and out-states, due to the effect of curved spacetime.
Thus if the solutions are adiabatic, no such mixing occurs and
thus no particle production occurs.

In order to have a concrete definition of the vacuum we need
adiabatic solutions for Eq. (\ref{mainequationperturb}), so we
employ a WKB recursion method of solving it, with solutions of the
form,
\begin{equation}\label{mainsolution}
\chi_k(\eta)=\frac{1}{2\omega(\eta)}e^{\pm
\int^{\eta}\omega(\eta)}\, ,
\end{equation}
where $\omega(\eta)$ is evaluated using a recursion formula, and
the leading order expression is,
\begin{equation}\label{recursionmain}
\omega(\eta)^2=\omega_k(\eta)\left(1+\delta_2(\eta)\omega_k(\eta)^{-2}+...
\right)\, ,
\end{equation}
where $\delta_2(\eta)$ is easily evaluated by using a recursion
series for $\omega(\eta)$, and it reads,
\begin{equation}\label{delta2}
\delta_2(\eta)=\frac{3}{4}\left(\frac{\omega_k'(\eta)}{\omega_k(\eta)}\right)^2-\frac{1}{2}\frac{\omega_k''(\eta)}{\omega_k(\eta)}\,
,
\end{equation}
where the ``prime'' denotes differentiation with respect to the
conformal time. The adiabaticity condition is that the leading
order term $\delta_2(\eta)\omega_k(\eta)^{-2}$, must satisfy,
\begin{equation}\label{adiabaticitycondition}
|\left|\delta_2(\eta)\omega_k(\eta)^{-2}\right|\ll 1\, .
\end{equation}
Now let us make contact with specific cosmological theories, so we
consider EGB theories in a flat FRW spacetime. In order to
quantify optimally the problem at hand, so in the next section we
recall the essential features of EGB inflationary theories.

\subsection{Overview of Viable EGB Inflationary Theories}

Viable EGB inflationary theories have many challenges to face,
since the theory must be simultaneously compatible with the Planck
data, and alternatively the ACT data, but also these theories must
be compatible with the GW170817 even which put serious constraints
on the propagation speed of the tensor perturbations. For
reference, the Planck data \cite{Planck:2018jri} and BICEP data
\cite{BICEP:2021xfz} constrain the spectral index of the scalar
perturbations as follows,
\begin{equation}\label{planck2018}
\centering n_{\mathcal{S}}=0.9649\pm0.0042,\,\,\, r<0.036\, .
\end{equation}
The Planck data are the benchmark in inflationary cosmology, at
least currently, but there are also the ACT data
\cite{ACT:2025fju,ACT:2025tim} which are in $2\sigma$ tension with
the Planck data, since the scalar spectral index was found to be
constrained as follows,
\begin{equation}\label{act}
n_{\mathcal{S}}=0.9743 \pm 0.0034\, .
\end{equation}
This tension has attracted the attention of theoretical
cosmologists in the field, and a considerable number of works has
been published, aiming to model inflationary models that are
compatible with the ACT data, see
\cite{Kallosh:2025rni,Gao:2025onc,Liu:2025qca,Yogesh:2025wak,Yi:2025dms,Peng:2025bws,Yin:2025rrs,Byrnes:2025kit,
Wolf:2025ecy,Aoki:2025wld,Gao:2025viy,Zahoor:2025nuq,Ferreira:2025lrd,Mohammadi:2025gbu,Choudhury:2025vso,
Odintsov:2025wai,Q:2025ycf,Zhu:2025twm,Kouniatalis:2025orn,Hai:2025wvs,Dioguardi:2025vci,Yuennan:2025kde,
Kuralkar:2025zxr,Kuralkar:2025hoz,Modak:2025bjv,
Aoki:2025ywt,Ahghari:2025hfy,McDonough:2025lzo,Chakraborty:2025wqn,NooriGashti:2025gug,Yuennan:2025mlg,
Deb:2025gtk,Afshar:2025ndm,Ellis:2025zrf,Iacconi:2025odq,Yuennan:2025tyx,Wang:2025cpp,Qiu:2025uot,Wang:2025dbj,Asaka:2015vza,Oikonomou:2025htz,Choudhury:2025hnu,Singh:2025uyr,Kim:2025dyi,Peng:2026ofs,newref,Odintsov:2026dss}
and references therein. Let us note that the ACT data are not
considered solely, but in conjunction with the DESI data
\cite{DESI:2024uvr}, and the DESI data have really stirred things
up in theoretical cosmology
\cite{Lee:2025pzo,Ozulker:2025ehg,Kessler:2025kju,Nojiri:2025low,Vagnozzi:2019ezj},
which added new mysteries in the already mysterious basis of
cosmology
\cite{Pedrotti:2024kpn,Jiang:2024xnu,Vagnozzi:2023nrq,Adil:2023exv,Bernui:2023byc,Gariazzo:2021qtg},
see also Ref. \cite{Sabogal:2026qvy} for fresh insights in the
field of inflationary cosmology and its interconnection to the
Hubble tension. So regarding the inflationary viability, we shall
consider both the benchmark Planck data and the ACT data and of
course GW170817 compatible theories only.

Let us return to EGB inflationary theories, and we present in
brief the formalism of EGB inflationary theories. The action of
EGB theories is,
\begin{equation}
\label{action} \centering
S=\int{d^4x\sqrt{-g}\left(\frac{R}{2\kappa^2}-\frac{1}{2}\partial_{\mu}\phi\partial^{\mu}\phi-V(\phi)-\frac{1}{2}\xi(\phi)\mathcal{G}\right)}\,
,
\end{equation}
with $R$ denoting the Ricci scalar, $\kappa=\frac{1}{M_p}$ with
$M_p$ being the reduced Planck mass, and in addition $\mathcal{G}$
is the four dimensional Gauss-Bonnet invariant, which is expressed
in terms of the Ricci scalar, the Ricci tensor $R_{\alpha\beta}$
and the Riemann tensor $R_{\alpha\beta\gamma\delta}$ in the
following way
$\mathcal{G}=R^2-4R_{\alpha\beta}R^{\alpha\beta}+R_{\alpha\beta\gamma\delta}R^{\alpha\beta\gamma\delta}$.
By assuming that the scalar field has no space-dependence, by
varying the action (\ref{action}) with respect to the scalar field
and the metric, we obtain the following field equations,
\begin{equation}
\label{motion1} \centering
\frac{3H^2}{\kappa^2}=\frac{1}{2}\dot\phi^2+V+12 \dot\xi H^3\, ,
\end{equation}
\begin{equation}
\label{motion2} \centering \frac{2\dot
H}{\kappa^2}=-\dot\phi^2+4\ddot\xi H^2+8\dot\xi H\dot H-4\dot\xi
H^3\, ,
\end{equation}
\begin{equation}
\label{motion3} \centering \ddot\phi+3H\dot\phi+V'+12 \xi'H^2(\dot
H+H^2)=0\, .
\end{equation}
The inflationary era occurs when $\dot{H}\ll H^2$, and also it is
mandatory to have a slow-roll inflationary era in order for the
inflationary regime to last long enough in order to solve the
horizon and flatness problems. So we need to require the following
slow-roll constraints
\begin{equation}\label{slowrollhubble}
\dot{H}\ll H^2,\,\,\ \frac{\dot\phi^2}{2} \ll V,\,\,\,\ddot\phi\ll
3 H\dot\phi\, .
\end{equation}
The speed of the tensor perturbations of the EGB is equal to
\cite{Hwang:2005hb},
\begin{equation}
\label{GW} \centering c_T^2=1-\frac{Q_f}{2Q_t}\, ,
\end{equation}
with the functions $Q_f$, $F$ and $Q_b$ being equal to $Q_f=8
(\ddot\xi-H\dot\xi)$, $Q_t=F+\frac{Q_b}{2}$,
$F=\frac{1}{\kappa^2}$ and in addition $Q_b=-8 \dot\xi H$. The
inflationary phenomenology can be derived by using the slow-roll
indices, and the observational indices are expressed in terms of
the slow-roll indices. The slow-roll indices for EGB theories have
the general form \cite{Hwang:2005hb},
\begin{align}
\centering \label{indices} \epsilon_1&=-\frac{\dot
H}{H^2}&\epsilon_2&=\frac{\ddot\phi}{H\dot\phi}&\epsilon_3&=0&\epsilon_4&=\frac{\dot
E}{2HE}&\epsilon_5&=\frac{Q_a}{2HQ_t}&\epsilon_6&=\frac{\dot
Q_t}{2HQ_t}\, ,
\end{align}
with $Q_a=-4\dot{\xi} H^2 $, $Q_b=-8\dot{\xi} H$,
$E=\frac{1}{(\kappa\dot\phi)^2}\left(
\dot\phi^2+\frac{3Q_a^2}{2Q_t}+Q_c\right)$, $Q_c=0$, $Q_d=0$,
$Q_e=-16 \dot{\xi} \dot{H}$, $Q_f=8\left(\ddot{\xi}-\dot{\xi}H
\right)$ and in addition $Q_t=\frac{1}{\kappa^2}+\frac{Q_b}{2}$.
The observational indices of inflation that are severely
constrained by the CMB experiments are the spectral index of the
primordial scalar perturbations, which is,
\begin{equation}\label{spectralindex}
n_{\mathcal{S}}=1+\frac{2 (-2
\epsilon_1-\epsilon_2-\epsilon_4)}{1-\epsilon_1}\, ,
\end{equation}
and also the tensor-to-scalar ratio which is equal to,
\begin{equation}\label{tensortoscalar}
r=\left |\frac{16 \left(c_A^3 \left(\epsilon_1-\frac{1}{4} \kappa
^2 \left(\frac{2
Q_c+Q_d}{H^2}-\frac{Q_e}{H}+Q_f\right)\right)\right)}{c_T^3
\left(\frac{\kappa ^2 Q_b}{2}+1\right)}\right |\, ,
\end{equation}
where $c_A$ is the sound speed of the scalar perturbations, the
exact form of which is,
\begin{equation}\label{soundspeed}
c_A=\sqrt{\frac{\frac{Q_a Q_e}{\frac{2}{\kappa ^2}+Q_b}+Q_f
\left(\frac{Q_a}{\frac{2}{\kappa
^2}+Q_b}\right)^2+Q_d}{\dot{\phi}^2+\frac{3 Q_a^2}{\frac{2}{\kappa
^2}+Q_b}+Q_c}+1}\, .
\end{equation}
This is an important quantity that will determine the adiabatic
perturbations. Also, the propagation speed of the tensor
perturbations, which is essentially the primordial gravitational
wave propagation speed, appears in Eq. (\ref{GW}) and this
quantity will also determine the adiabaticity conditions for the
perturbations. The spectral index of the tensor perturbations has
the following form,
\begin{equation}\label{tensorspectralindex}
n_{\mathcal{T}}=-2\frac{\left( \epsilon_1+\epsilon_6
\right)}{1-\epsilon_1}\, .
\end{equation}
The last observational quantity which is necessary for EGB
inflationary phenomenology is the amplitude of the scalar
perturbations $\mathcal{P}_{\zeta}(k_*)$ which is constrained by
the Planck data \cite{Planck:2018jri} as follows
$\mathcal{P}_{\zeta}(k_*)=2.196^{+0.051}_{-0.06}\times 10^{-9}$.
The definition of the amplitude of the scalar perturbations is,
\begin{equation}\label{definitionofscalaramplitude}
\mathcal{P}_{\zeta}(k_*)=\frac{k_*^3}{2\pi^2}P_{\zeta}(k_*)\, ,
\end{equation}
and it must be evaluated at first horizon crossing at the
beginning of inflation, with $k_*$ being the CMB pivot scale. In
the case of EGB theories, the amplitude of the scalar
perturbations $\mathcal{P}_{\zeta}(k)$ can be written in terms of
the slow-roll parameters as follows \cite{Hwang:2005hb},
\begin{equation}\label{powerspectrumscalaramplitude}
\mathcal{P}_{\zeta}(k)=\left(\frac{k \left((-2
\epsilon_1-\epsilon_2-\epsilon_4) \left(0.57\, +\log \left(\left|k
\eta \right| \right)-2+\log (2)\right)-\epsilon_1+1\right)}{(2 \pi
) \left(z c_A^{\frac{4-n_{\mathcal{S}}}{2}}\right)}\right)^2\, ,
\end{equation}
with $z=\frac{a \dot{\phi} \sqrt{\frac{E(\phi )}{\frac{1}{\kappa
^2}}}}{H (\epsilon_5+1)}$, with
$\eta=-\frac{1}{aH}\frac{1}{-\epsilon_1+1}$,  and all the
quantities appearing above must be calculated at the first horizon
crossing.

Now there are two classes of GW170817 compatible EGB theories,
which produce a viable inflationary regime. These were developed
in Refs. \cite{Odintsov:2020sqy,Oikonomou:2021kql} where a
constraint satisfied by the non-minimal Gauss-Bonnet coupling
$\xi(\phi)$ is satisfied, and in Ref. \cite{Oikonomou:2024etl}
where the resulting theory produces a gravitational wave speed
compatible with the GW170817 constraints. We shall call the former
class of EGB theories of Refs.
\cite{Odintsov:2020sqy,Oikonomou:2021kql} constrained EGB gravity,
and the theory developed in \cite{Oikonomou:2024etl},
unconstrained EGB gravity.

\subsection{Conditions for Adiabatic Perturbations in EGB Gravity}

Let us now derive the Mukhanov-Sasaki equation for the scalar and
tensor perturbations of EGB gravity. Our aim is to reveal the
actual form of Eq. (\ref{mainequationperturb}) for the
cosmological perturbations of EGB and to see at first hand the
conditions that adiabaticity imposes on the minimal Gauss-Bonnet
coupling. Then we can work out numerically and in some cases
analytically the conditions of adiabaticity and see when the
theory at hand produces non-adiabatic perturbations. Finally we
discuss qualitatively later on the effects of adiabaticity
violation.

Let us derive the perturbation equations for EGB gravity based on
the presentation of \cite{Hwang:2005hb}. The metric perturbation
of the flat FRW spacetime is,
 \bea
   & & d s^2 = - a^2 \left( 1 + 2 \alpha \right) d \eta^2
       - 2 a^2 \beta_{,\alpha} d \eta d x^\alpha
       + a^2 \left( g^{(3)}_{\alpha\beta}
       + 2 \varphi g^{(3)}_{\alpha\beta}
       + 2 \gamma_{,\alpha|\beta}
       + 2 C_{\alpha\beta} \right) d x^\alpha d x^\beta,
   \label{metric}
\eea where $a(\eta)$ is the scale factor expressed in terms of the
conformal time $\eta$. The parameters $\alpha$, $\beta$, $\gamma$
and $\varphi$ quantify the scalar spacetime-dependent
perturbations, and also $C_{\alpha\beta}$ stands for the tensor
perturbation and it is a transverse and trace-free tensor. Also
$g^{(3)}_{\alpha\beta}$ denotes the metric of the comoving
three-space section corresponding to the FRW metric, \bea
   g^{(3)}_{\alpha\beta} d x^\alpha d x^\beta
   &=& {1 \over \left( 1 + \bar r^2 \right)^2}
       \left( d x^2 + d y^2 + d z^2 \right)\,. \eea
We can express the kinematic quantities in the normal frame in the
following way, \bea
   & & \theta = 3 H , \quad
       \sigma_{\alpha\beta}
       = \chi_{,\alpha|\beta}
       - {1 \over 3} g_{\alpha\beta}^{(3)} \Delta \chi
       + a^2 \dot C^{(t)}_{\alpha\beta}, \quad
       a_\alpha = \alpha_{,\alpha}, \quad
       R^{(h)} = {1 \over a^2} \left[ 6 K - 4 \Delta
       \varphi \right],
   \label{kinematic-quantities}
\eea where \bea
   & & \chi \equiv a \left( \beta + a \dot \gamma \right),
   \label{chi-def}
\eea and also $\Delta$ denotes the Laplacian operator of
$g^{(3)}_{\alpha\beta}$. In addition $\theta$ denotes the
expansion scalar, also $\sigma_{ab}$ denotes the shear tensor, and
in addition $a_a$ stands for the acceleration vector. By
performing the gauge transformation, $\hat x^a \equiv x^a + \tilde
\xi^a (x^e)$ we obtain the following expressions for the
perturbation variables, \bea
   & & \hat \alpha = \alpha - \dot \xi^t, \quad
       \hat \beta = \beta - {1 \over a} \xi^t
       + a \left( {\xi \over a} \right)^\cdot, \quad
       \hat \gamma = \gamma - {1 \over a} \xi, \quad
       \hat \varphi = \varphi - H \xi^t, \quad
       \hat \chi = \chi - \xi^t, \quad
       \hat \kappa = \kappa
       + \left( 3 \dot H + {\Delta \over a^2} \right) \xi^t,
   \nonumber \\
   & & \delta \hat \mu = \delta \mu - \dot \mu \xi^t, \quad
       \delta \hat p = \delta p - \dot p \xi^t, \quad
       \hat v = v - {1 \over a} \xi^t, \quad
       \hat \Pi = \Pi, \quad
       \delta \hat \phi = \delta \phi - \dot \phi \xi^t; \quad
       \hat C_{\alpha\beta} = C_{\alpha\beta}, \quad
       \hat \Pi_{\alpha\beta}^{({t})} = \Pi_{\alpha\beta}^{({t})},
   \label{GT}
\eea with $\xi^0 \equiv {1 \over a} \xi^t$ and moreover
$\xi_\alpha \equiv \xi_{,\alpha}$. In addition, $\bar \phi$ and
$\delta \phi$ stand for the background and also the perturbation
part of the scalar field $\phi({\bf x}, t)$. We shall use the
following gauge-invariant variables,
\begin{align}
 & \varphi_\chi \equiv \varphi - H \chi, \,\,\,
       \varphi_v \equiv \varphi - a H v, \,\,\,
       \delta_v \equiv \delta - a {\dot \mu \over \mu} v, \,\,\,
       \delta \phi_\varphi \equiv \delta \phi
       - {\dot \phi \over H} \varphi
       \equiv - {\dot \phi \over H} \varphi_{\delta \phi}, \\
       \notag &
       v_\chi \equiv v - {1 \over a} \chi
       \equiv - {1 \over a} \chi_v,
   \label{GI-variables}
\end{align}
with $\delta \equiv \delta \mu / \mu$. In addition, we shall use
the perturbation variable definitions that follow, \bea
   & & \Phi \equiv \varphi_{\delta \phi}, \quad
       \Psi \equiv \varphi_\chi + {\dot F + Q_a \over 2F + Q_b}
       {\delta F_\chi \over \dot F}.
   \label{Psi-def-string}
\eea and in addition the sound wave speed is, \bea
   & & c_A^2 = 1 + { Q_d + {\dot F + Q_a \over 2F + Q_b} Q_e
       + \left( {\dot F + Q_a \over 2F + Q_b} \right)^2 Q_f
       \over \omega \dot \phi^2 + 3 {(\dot F + Q_a)^2 \over 2 F + Q_b} + Q_c }.
\eea The scalar-type perturbation equations for EGB,   are written
in the following way, \bea
   & & \dot \Phi = 2 x_1 {\Delta \over a^2} \Psi,
   \label{dot-Phi-eq} \\
   & & {1 \over x_2} \left( x_2 \Psi \right)^\cdot = {1 \over 2} x_3
   \Phi,
   \label{dot-Psi-eq}
\eea with,
\begin{equation}\label{x1}
x_1={ \left( H + {\dot F + Q_a \over 2 F + Q_b} \right)
         \left( F + {1 \over 2} Q_b \right)
         \over \omega \dot \phi^2
         + 3 {(\dot F + Q_a)^2 \over 2 F + Q_b} + Q_c }\, ,
\end{equation}
\begin{equation}\label{x2}
 x_2={a (F + {1 \over 2} Q_b) \over
       H + {\dot F + Q_a \over 2 F + Q_b}}\, ,
\end{equation}
\begin{equation}\label{x3}
x_3={1 \over \left( H + {\dot F + Q_a \over 2 F + Q_b} \right)
         \left( F + {1 \over 2} Q_b \right)} x_4\, ,
\end{equation}
\begin{equation}\label{x24}
x_4=\dot \phi^2 + 3 {(\dot F + Q_a)^2 \over 2 F + Q_b}
  + Q_c + Q_d + {\dot F + Q_a \over 2F + Q_b} Q_e
  + \left( {\dot F + Q_a \over 2F + Q_b} \right)^2 Q_f\, ,
\end{equation}
and $F=\frac{1}{\kappa^2}$. Upon normalizing, and redefining the
variables in the following way, \bea
   & & z= \frac{a\dot{\phi}}{H}, \quad
    \quad
       \tilde v \equiv z \Phi, \quad
       u \equiv \frac{1}{\kappa^2}\frac{a}{H} {1 \over z} \Psi,
   \label{u-v-def}
\eea we get the perturbation equations,
 \bea
   & & \tilde v^{\prime\prime}
       - \left( c_A^2 \Delta + {z^{\prime\prime} \over z} \right) \tilde v
       = a^2 z \left[ {1 \over a z^2} \left( a z^2 \dot \Phi \right)^\cdot
       - c_A^2 {\Delta \over a^2} \Phi \right] = 0,
   \label{v-eq} \\
   & & u^{\prime\prime} - \left[ c_A^2 \Delta
       + {(1/\bar z)^{\prime\prime} \over (1/\bar z)} \right] u
       = {a^2 x_2 \over \bar z}
       \left\{ {\bar z^2 \over a x_2} \left[ {a \over \bar z^2}
       \left( x_2 \Psi \right)^\cdot \right]^\cdot
       - c_A^2 {\Delta \over a^2} \Psi \right\} = 0\, ,
   \label{u-eq}
\eea and $c_A$ is the sound wave speed of the fluctuating fluid or
the fluctuating field and of the excited metric. Regarding the
tensor mode, by using,
\bea
   & & z_t \equiv a \sqrt{Q_t}, \quad
       v_t \equiv z_t \Phi,
   \label{z-v-def-GW}
\eea with $\Phi = C_{\alpha\beta}$ or $h_{\ell {\bf k}}$, we
obtain, \bea
   & & v_t^{\prime\prime}
       - \left( c_T^2 \Delta + {z_t^{\prime\prime} \over z_t} \right) v_t
       = a^2 z_t \left[
       {1 \over a z_t^2} \left( a z_t^2 \dot \Phi \right)^\cdot
       - c_T^2 {\Delta \over a^2} \Phi \right]
       = 0.
   \label{v-eq-GW}
\eea The two differential equations (\ref{v-eq} and \ref{v-eq-GW})
are the Mukhanov-Sasaki perturbation equations for EGB gravity and
specifically the scalar and tensor perturbation equations
respectively.

Now we can proceed by specifying the exact form of
$\omega_k(\eta)$ in Eq. (\ref{mainequationperturb}) for the scalar
and tensor perturbations of EGB gravity. Let us start with the
scalar perturbations, so the second order perturbed Mukhanov
action for the scalar perturbation $\tilde{v}$ is,
\begin{equation}\label{pertubredmukhanovaction}
\delta S^2=\frac{1}{2}\int \left( \tilde{v}''^2-c_A^2
\tilde{v}^{,\mu}\tilde{v}_{,\mu}+\frac{z''}{z}\tilde{v}^2
\right)\mathrm{d}x^3\mathrm{d}\nu\, ,
\end{equation}
so by using the definition above $\tilde{v}=z\Phi$, we get the
perturbation equation for the scalar perturbations,
\begin{equation}\label{scalarperturbationseqn}
\tilde v^{\prime\prime}
       - \left(- c_A^2 k^2 + {z^{\prime\prime} \over z} \right) \tilde
       v=0\, ,
\end{equation}
and expanding $\tilde{v}$ in Fourier modes, we obtain,
\begin{equation}\label{fourierscalarperturbations}
\tilde{v}=\int \frac{\mathrm{d}k^3}{(2\pi)^3}\left(a_k v_k e^{i k
x}+a_k^{\dag}v_k^*e^{-i k x}\right)\, .
\end{equation}
Then each Fourier mode $v_k$ of the scalar perturbation
$\tilde{v}$ satisfies the following evolution equation,
\begin{equation}\label{scalarperturbationseqnfinal}
v_k^{\prime\prime}
       - \left(- c_A^2 k^2 + {z^{\prime\prime} \over z} \right)
       v_k=0\, .
\end{equation}
The differential equation (\ref{scalarperturbationseqnfinal}) is
identical to Eq. (\ref{mainequationperturb}) so we can identify,
\begin{equation}\label{identifomega1}
\omega_k(\eta)=c_A^2k^2-\frac{z^{\prime \prime}}{z}\, .
\end{equation}
Now in view of Eq. (\ref{recursionmain}) and also (\ref{delta2}),
the adiabaticity condition (\ref{adiabaticitycondition}) at
leading order reads,
\begin{equation}\label{adiabaticity1}
\left|\frac{\omega_k^{\prime}}{\omega_k^2}\right|\ll 1\, .
\end{equation}
Considering the subhorizon modes, we have $\omega_k\sim c_A k$,
thus the adiabaticity condition becomes,
\begin{equation}\label{adiabaticity2}
\left|\frac{c_A^{\prime}}{c_A^2k}\right|\ll 1\, ,
\end{equation}
or in terms of the cosmic time,
\begin{equation}\label{adiabaticity3}
\left|\frac{a \dot{c}_A}{c_A^2k}\right|\ll 1\, .
\end{equation}
By introducing the Hubble rate $H$ at both hands of the above
equation, the condition (\ref{adiabaticity3}) can easily be
written as follows,
\begin{equation}\label{adiabaticity4}
\left|\frac{\dot{c}_A}{c_A H}\right|\ll \frac{k c_A}{a H}\, ,
\end{equation}
and since we are considering subhorizon modes, the condition
(\ref{adiabaticity4}) reads,
\begin{equation}\label{adiabaticityscalarfinalsoundspeed}
\left|\frac{\dot{c}_A}{c_A H}\right|\ll 1\, .
\end{equation}
Now the violation of adiabaticity occurs near the horizon crossing
of the subhorizon modes, when $c_A k\sim a H$, so when,
\begin{equation}\label{nonadiabatictycondition}
 \left|\frac{\dot{c}_A}{c_A H}\right|\sim 1\, ,
\end{equation}
the adiabaticity is violated. In the same way, the adiabaticity
criterion for the tensor perturbations is,
\begin{equation}\label{expressionfinaloizedtensor}
\left|\frac{\dot{c}_T}{c_T H}\right|\ll 1\, .
\end{equation}

Now suppose that the adiabaticity criteria are violated, so we
shall provide some general formulas for EGB gravity, which will
cover the adiabaticity investigation to be presented in the next
sections. Suppose we start from an adiabatic vacuum,
\begin{equation}\label{bdv}
v_k=\frac{1}{\sqrt{\omega_k}}e^{-i\int^{\eta}\omega_k\mathrm{d}\eta}\,
,
\end{equation}
so this is the ``in'' cosmological state asymptotically in the
past at the beginning of inflation. The cosmological ``out'' state
is,
\begin{equation}\label{outstate}
v_k=\frac{1}{\sqrt{\omega_k}}\left(
a_k(\eta)e^{-i\int^{\eta}\omega_k\mathrm{d}\eta}+\beta_k(\eta)e^{i\int^{\eta}\omega_k\mathrm{d}\eta}\right)\,
,
\end{equation}
hence positive and negative solutions mix in the ``out state''.
The problematic term, that will cause particle production if the
adiabaticity condition is violated, is the negative frequency term
that contains the Bogoliubov coefficient $\beta_k(\eta)$, which is
equal to,
\begin{equation}\label{betak}
\beta_k(\eta)=\int_{-\infty}^{\eta}\mathrm{d}\tau
\frac{\omega_k^{\prime}(\tau)}{2\omega_k(\tau)}e^{-2 i
\int^{\eta}\omega_k(\tau)\mathrm{d}\tau}\, ,
\end{equation}
which stems from the evolution equation,
\begin{equation}\label{betakevolution}
\beta_k'(\eta)=\frac{\omega^{\prime}_k(\eta)}{2\omega_k(\eta)}e^{-2
i \int^{\eta}\omega(\tau)\mathrm{d}\tau}\, .
\end{equation}
Eq. (\ref{betak}) is easily converted to contain the cosmic time,
\begin{equation}\label{betakcosmictime}
\beta_k(t)=\int_{-\infty}^{t}\mathrm{d}t
\frac{\dot{\omega}_k(t)}{2\omega_k(t)}e^{-2 i
\int^{t}\frac{\omega_k(t)}{a(t)}\mathrm{d}t}\, .
\end{equation}
In the following sections we shall investigate the implications of
the adiabaticity conditions developed in this section to two
mainstream EGB inflationary approaches appearing in the
literature.

\section{GW170817-compatible Unconstrained EGB Theories are Also Fully Adiabatic}

In this section we examine the adiabaticity of the inflationary
perturbations for the unconstrained EGB inflationary theories that
respect the GW170817 constraint (\ref{constraintGW170817}). For
the phenomenology of this section, and the assignment of numerical
values in the free parameters, we shall use the Planck units
system,
\begin{equation}\label{planckunits}
\kappa^2=\frac{1}{8\pi G}=\frac{1}{M_P^2}=1\, ,
\end{equation}
however for the expressions we shall keep the Planck related
parameter $\kappa=\frac{1}{M_p}$, in order to be formal. Only in
the numerical analysis we shall use Planck units for simplicity.

We shall examine if the adiabaticity constraints
(\ref{adiabaticityscalarfinalsoundspeed}) are satisfied for the
inflationary viable theories developed in Ref.
\cite{Oikonomou:2024etl}. The constraints imposed on the EGB
inflationary theories are indeed large in number and too
restrictive but the models of this section succeed to pass all the
phenomenological tests, and even the adiabaticity constraints. Let
us recall the formalism before we proceed to the examination of
the models, which will be numerical for the whole duration of the
inflationary regime. Recall that the constraint
(\ref{constraintGW170817}) requires the gravitational wave speed
of Eq. (\ref{GW}) to have significantly small contributions
related to the Gauss-Bonnet coupling function. From the
gravitational wave speed functional form (\ref{GW}), we can see
that the constraint (\ref{constraintGW170817}) can be satisfied
if,
\begin{equation}\label{actualgw170817constraints}
\kappa^2\dot{\xi}H\ll 1,\,\,\,\kappa^2\ddot{\xi}\ll 1\, .
\end{equation}
These two constraints have to be satisfied independently from each
other, and note that these two constraints are fundamental and are
not chosen for the sake of analyticity. These constraints realize
the requirement that $c_T^2\sim 1$, and in this context, the
Gauss-Bonnet coupling is free to chose, in contrast to the
approach adopted in Refs.
\cite{Oikonomou:2021kql,Oikonomou:2022xoq,Odintsov:2020sqy} which
will be discussed in the next section. Now in the present context,
due to the assumptions (\ref{actualgw170817constraints}) and also
taking into account the slow-roll assumptions, the Friedmann
equation reads,
\begin{equation}
\label{motion5} \centering H^2\simeq\frac{\kappa^2V}{3}\, ,
\end{equation}
and by making the effective theory motivated assumptions,
\begin{equation}\label{additionalconstraints}
\kappa^2\dot{\xi}H^3\ll
\kappa^2\dot{\phi}^2,\,\,\,\kappa^2\ddot{\xi}H^2\ll
\kappa^2\dot{\phi}^2\, ,
\end{equation}
the Raychaudhuri equation reads,
\begin{equation}
\label{motion6} \centering \dot H\simeq-\frac{1}{2}\kappa^2
\dot\phi^2\, ,
\end{equation}
while the modified Klein-Gordon equation becomes,
\begin{equation}
\label{motion8} \centering \dot\phi\simeq
-\frac{12\xi'(\phi)H^4+V'}{3H}\, .
\end{equation}
Now, having at hand the structural field equations at an effective
field theory leading order level, we can proceed to the
phenomenology of the models of this section. The $e$-foldings
number is,
\begin{equation}
\label{efolds} \centering
N=\int_{t_i}^{t_f}{Hdt}=\int_{\phi_i}^{\phi_f}\frac{H}{\dot{\phi}}d\phi\,
,
\end{equation}
with $\phi_i$ and $\phi_f$ being the scalar field values at first
horizon crossing, and at the end of inflation respectively. Taking
into account (\ref{efolds}) and (\ref{motion8}), the $e$-foldings
number (\ref{efolds}) takes the following form,
\begin{equation}
\label{efolds1} \centering
N=\int_{\phi_i}^{\phi_f}\frac{2H^2}{12\xi'H^4+V'}d\phi\, ,
\end{equation}
which in view of the Friedmann equation (\ref{motion5}) is finally
written as follows,
\begin{equation}
\label{efoldsuncostrained} \centering
N=\int_{\phi_f}^{\phi_i}\frac{\kappa ^2 V(\phi )}{V'(\phi
)+\frac{4}{3} \kappa ^4 V(\phi )^2 \xi '(\phi )}d\phi\, .
\end{equation}
At this point, we can express the  first slow-roll index in terms
of the Gauss-Bonnet scalar coupling and the scalar potential, so
combining Eqs. (\ref{motion5}), (\ref{motion6}) and
(\ref{motion8}), we have,
\begin{equation}\label{epsilon1analytic}
\epsilon_1=\frac{4}{3} \kappa ^2 \xi '(\phi ) V'(\phi
)+\frac{V'(\phi )^2}{2 \kappa ^2 V(\phi )^2}+\frac{8}{9} \kappa ^6
V(\phi )^2 \xi '(\phi )^2\, .
\end{equation}
In order to have full analytic command of the inflationary
phenomenology, we may choose the Gauss-Bonnet scalar coupling
function as follows,
\begin{equation}\label{couplingfunctionchoices3}
\xi'(\phi)=\frac{\lambda  V'(\phi )}{V(\phi )^2}\, ,
\end{equation}
so the gravitational wave speed becomes,
\begin{equation}\label{gwspeedclass2}
c_T^2=\frac{-8 \lambda  (4 \lambda +3)^2 V(\phi ) V'(\phi )^2
V''(\phi )+10 \lambda  (4 \lambda +3)^2 V'(\phi )^4+27 V(\phi
)^4}{3 V(\phi )^2 \left(4 \lambda  (4 \lambda +3) V'(\phi )^2+9
V(\phi )^2\right)}\, ,
\end{equation}
and the first slow-roll index takes the form,
\begin{equation}\label{epsilon1formclass}
\epsilon_1=\frac{(4 \lambda +3)^2 V'(\phi )^2}{18 V(\phi )^2}\, ,
\end{equation}
while the $e$-foldings number takes the form,
\begin{equation}\label{finalinitialefoldings}
N=\int_{\phi_f}^{\phi_i} \frac{V(\phi )}{\frac{4}{3} \lambda
V'(\phi )+V'(\phi )} \mathrm{d}\phi\, .
\end{equation}
An interesting aspect of the models
(\ref{couplingfunctionchoices3}) is that the coupling term
increases when the potential decreases during inflation
\cite{vandeBruck:2016xvt}.

Having the formalism at hand, we can proceed to examining the
phenomenology of some models, and also examine numerically whether
the adiabaticity constraints are satisfied. Let us start with the
following model, in which the scalar potential is chosen as
follows,
\begin{equation}\label{viablepotentials}
V(\phi)=M \left(1-\frac{\delta }{\kappa  \phi }\right)\, ,
\end{equation}
and $\xi'(\phi)$ in this case is,
\begin{equation}\label{xiphi}
\xi'(\phi)=\frac{\delta  \kappa  \lambda }{M (\delta -\kappa  \phi
)^2}\, ,
\end{equation}
and the first slow-roll index is,
\begin{equation}\label{slowrollindexena}
\epsilon_1=\frac{\delta ^2 \left(4 \kappa ^4 \lambda
+3\right)^2}{18 \kappa ^2 \phi ^2 (\delta -\kappa  \phi )^2}\, ,
\end{equation}
while the $e$-foldings number is,
\begin{equation}\label{integraln}
N=-\frac{3 \kappa ^2 \left(\frac{\delta  \phi ^2}{2}-\frac{\kappa
\phi ^3}{3}\right)}{\delta  \left(4 \kappa ^4 \lambda
+3\right)}\Big{|}_{\phi_f}^{\phi_i}\, .
\end{equation}
This model can be compatible with the ACT data as we now evince,
for various choices of the free parameters of the model. For
example the compatibility with the ACT data comes for the choice
$N=60$ and $(\delta,\lambda,M)=(10^{-3},10^{-18},2.52\times
10^{-12})$, in which case we get $n_{\mathcal{S}}=0.97777$, $r=
0.0001574$ and the gravitational wave speed is
$|c_T^2-1|=1.28\times 10^{-23}$ and also the amplitude of the
scalar perturbations is $\mathcal{P}_{\zeta}(k_*)=2.196\times
10^{-9}$.
\begin{figure}
\centering
\includegraphics[width=18pc]{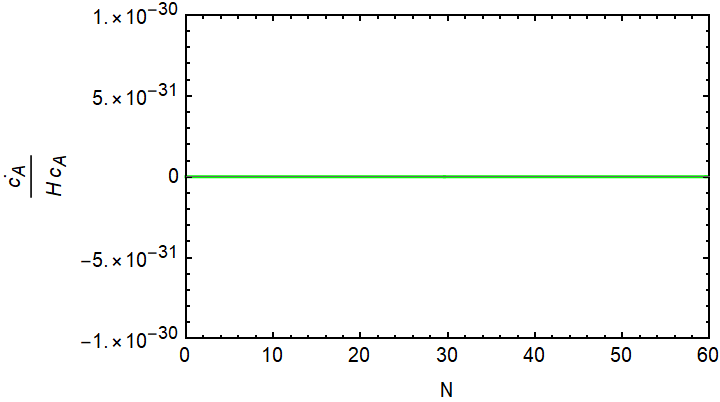}
\includegraphics[width=18pc]{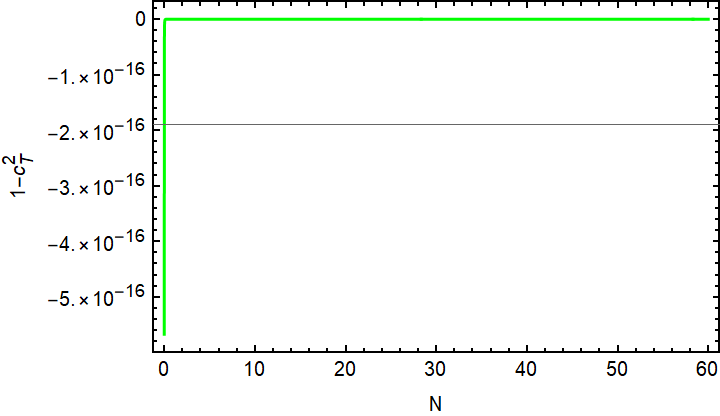}
\caption{The adiabaticity constraint $\frac{\dot{c}_A}{c_A H}$
(left plot) and the gravitational wave speed (right plot) as
functions of the $e$-foldings number for the model
(\ref{viablepotentials}).}\label{plot1}
\end{figure}
Also the adiabaticity constraint is satisfied for the whole
duration of the inflationary era. This can be seen in Fig.
\ref{plot1} where we present the behavior of the constraint
$\frac{\dot{c}_A}{c_A H}$ as a function of the $e$-foldings number
(left plot) and the gravitational wave speed as a function of the
$e$-foldings (right plot). As it can be seen, the resulting theory
is phenomenologically viable and it also produces adiabatic
perturbations, for the whole inflationary era. The numerical
values of $\frac{\dot{c}_A}{c_A H}$ are too small, so we quote
here the actual values at the end and the beginning of inflation
and these are $\frac{\dot{c}_A}{c_A
H}\Big{|}_{\phi_f}=-1.5130\times 10^{-33}$ and
$\frac{\dot{c}_A}{c_A H}\Big{|}_{\phi_i}=-5.257\times 10^{-48}$
respectively.

Now let us proceed with a model that is both ACT and Planck
compatible, in which case the scalar potential is chosen as
follows,
\begin{equation}\label{viablepotentials1}
V(\phi)=M \left(1-\frac{\delta }{\kappa  \phi ^2}\right)\, ,
\end{equation}
and $\xi'(\phi)$ in this case is,
\begin{equation}\label{xiphi1}
\xi'(\phi)=\frac{2 \delta  \kappa  \lambda  \phi }{M \left(\delta
-\kappa  \phi ^2\right)^2}\, ,
\end{equation}
and the first slow-roll index is,
\begin{equation}\label{slowrollindexena1}
\epsilon_1=\frac{2 \delta ^2 \left(4 \kappa ^4 \lambda
+3\right)^2}{9 \kappa ^2 \phi ^2 \left(\delta -\kappa  \phi
^2\right)^2}\, ,
\end{equation}
while the $e$-foldings number is,
\begin{equation}\label{integraln1}
N=-\frac{3 \kappa ^2 \left(\frac{\delta  \phi ^2}{2}-\frac{\kappa
\phi ^4}{4}\right)}{2 \delta  \left(4 \kappa ^4 \lambda
+3\right)}\Big{|}_{\phi_f}^{\phi_i}\, .
\end{equation}
This model can be compatible with both the ACT data and the Planck
data as we now evince, for various choices of the free parameters
of the model. For example the compatibility with the ACT data
comes for the choice $N=60$ and
$(\delta,\lambda,M)=(1.2,10^{-18},1.06\times 10^{-10})$, in which
case we get $n_{\mathcal{S}}=0.9750$, $r= 0.006329$ and the
gravitational wave speed is $|c_T^2-1|=5.1436\times 10^{-22}$ and
also the amplitude of the scalar perturbations is
$\mathcal{P}_{\zeta}(k_*)=2.196\times 10^{-9}$. With regard to the
compatibility of the model with respect to the Planck data, the
compatibility with the Planck data comes for the choice $N=50$ and
$(\delta,\lambda,M)=(10^{-3},10^{-20},1.06\times 10^{-10})$, in
which case we get $n_{\mathcal{S}}=0.96876$, $r= 0.0002683$ and
the gravitational wave speed is $|c_T^2-1|=2.0705\times 10^{-18}$
and also the amplitude of the scalar perturbations is
$\mathcal{P}_{\zeta}(k_*)=2.196\times 10^{-9}$.
\begin{figure}
\centering
\includegraphics[width=18pc]{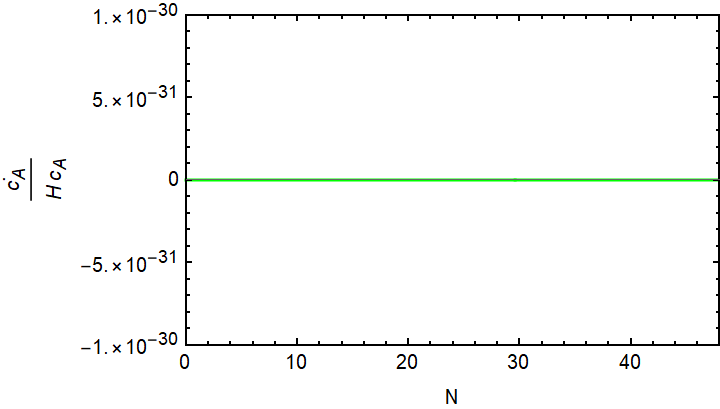}
\includegraphics[width=18pc]{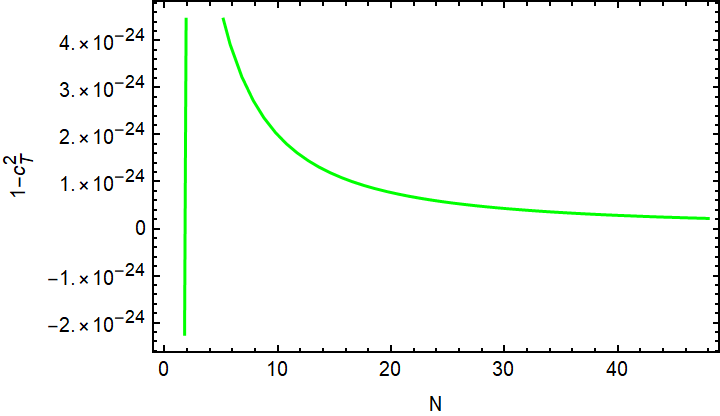}
\caption{The adiabaticity constraint $\frac{\dot{c}_A}{c_A H}$
(left plot) and the gravitational wave speed (right plot) as
functions of the $e$-foldings number for the model
(\ref{viablepotentials1}), taking into account the Planck data
compatibility of the model.}\label{plot2}
\end{figure}
Also the adiabaticity constraint is satisfied for the whole
duration of the inflationary era. This can be seen in Fig.
\ref{plot2} where we present the behavior of the constraint
$\frac{\dot{c}_A}{c_A H}$ as a function of the $e$-foldings number
(left plot) and the gravitational wave speed as a function of the
$e$-foldings (right plot). As it can be seen, the resulting theory
is phenomenologically viable and it also produces adiabatic
perturbations, for the whole inflationary era. The numerical
values of $\frac{\dot{c}_A}{c_A H}$ are too small, so we quote
here the actual values at the end and the beginning of inflation
and these are $\frac{\dot{c}_A}{c_A
H}\Big{|}_{\phi_f}=-5.52158\times 10^{-34}$ and
$\frac{\dot{c}_A}{c_A H}\Big{|}_{\phi_i}=-1.56149\times 10^{-47}$
respectively, using the Planck data compatibility. The same
viability applies for the ACT data too.

Now let us proceed with a model that is ACT compatible only, in
which case the scalar potential is chosen as follows,
\begin{equation}\label{viablepotentials2}
V(\phi)=M \left(1-\frac{\delta }{\kappa  \phi }\right)^2\, ,
\end{equation}
and $\xi'(\phi)$ in this case is,
\begin{equation}\label{xiphi2}
\xi'(\phi)=-\frac{2 \delta  \kappa ^2 \lambda  \phi }{M (\delta
-\kappa  \phi )^3}\, ,
\end{equation}
and the first slow-roll index is,
\begin{equation}\label{slowrollindexena2}
\epsilon_1=\frac{2 \delta ^2 \left(4 \kappa ^4 \lambda
+3\right)^2}{9 \kappa ^2 \phi ^2 (\delta -\kappa  \phi )^2}\, ,
\end{equation}
while the $e$-foldings number is,
\begin{equation}\label{integraln2}
N=-\frac{3 \kappa ^2 \left(\frac{\delta  \phi ^2}{2}-\frac{\kappa
\phi ^3}{3}\right)}{2 \delta  \left(4 \kappa ^4 \lambda
+3\right)}\Big{|}_{\phi_f}^{\phi_i}\, .
\end{equation}
This model can be compatible wit the ACT data only as we now
evince, for various choices of the free parameters of the model.
For example the compatibility with the ACT data comes for the
choice $N=60$ and $(\delta,\lambda,M)=(1.2,10^{-20},1.06\times
10^{-10})$, in which case we get $n_{\mathcal{S}}=0.97623$, $r=
0.02774$ and the gravitational wave speed is
$|c_T^2-1|=1.2748\times 10^{-19}$ and also the amplitude of the
scalar perturbations is $\mathcal{P}_{\zeta}(k_*)=2.196\times
10^{-9}$.
\begin{figure}
\centering
\includegraphics[width=18pc]{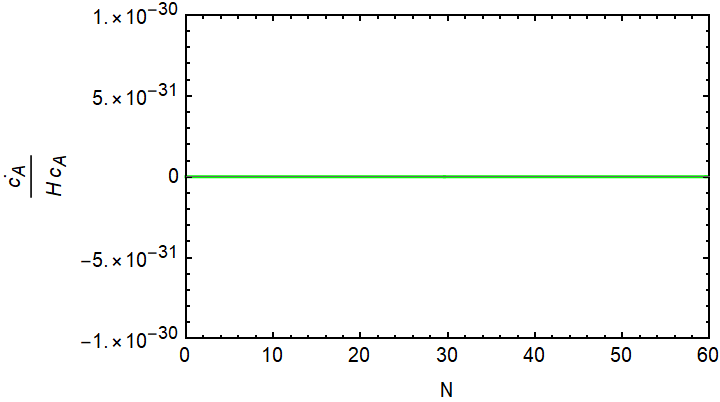}
\includegraphics[width=18pc]{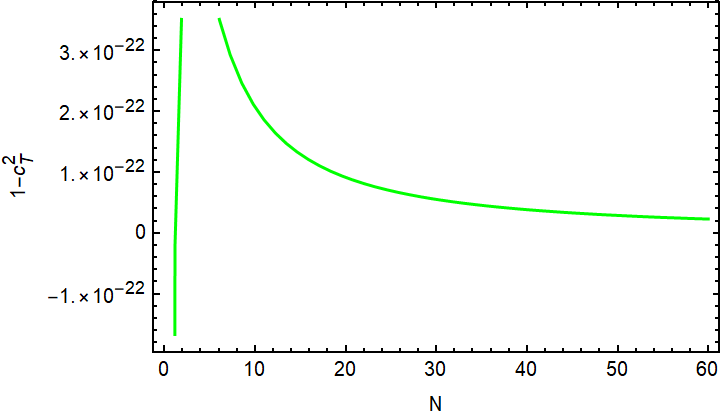}
\caption{The adiabaticity constraint $\frac{\dot{c}_A}{c_A H}$
(left plot) and the gravitational wave speed (right plot) as
functions of the $e$-foldings number for the model
(\ref{viablepotentials2}).}\label{plot3}
\end{figure}
Also the adiabaticity constraint is satisfied for the whole
duration of the inflationary era. This can be seen in Fig.
\ref{plot3} where we present the behavior of the constraint
$\frac{\dot{c}_A}{c_A H}$ as a function of the $e$-foldings number
(left plot) and the gravitational wave speed as a function of the
$e$-foldings (right plot). As it can be seen, the resulting theory
is phenomenologically viable and it also produces adiabatic
perturbations, for the whole inflationary era. The numerical
values of $\frac{\dot{c}_A}{c_A H}$ are too small, so we quote
here the actual values at the end and the beginning of inflation
and these are $\frac{\dot{c}_A}{c_A
H}\Big{|}_{\phi_f}=-3.39969\times 10^{-39}$ and
$\frac{\dot{c}_A}{c_A H}\Big{|}_{\phi_i}=-1.17676\times 10^{-47}$
respectively.

Now let us proceed with a model that is only ACT compatible, in
which case the scalar potential is chosen as follows,
\begin{equation}\label{viablepotentials3}
V(\phi)=M \sqrt{1-\frac{d}{\kappa  \phi }}\, ,
\end{equation}
and $\xi'(\phi)$ in this case is,
\begin{equation}\label{xiphi3}
\xi'(\phi)=\frac{d \lambda }{2 \kappa  M \phi ^2
\left(1-\frac{d}{\kappa  \phi }\right)^{3/2}}\, ,
\end{equation}
and the first slow-roll index is,
\begin{equation}\label{slowrollindexena3}
\epsilon_1=\frac{d^2 \left(4 \kappa ^4 \lambda +3\right)^2}{72
\kappa ^2 \phi ^2 (d-\kappa  \phi )^2}\, ,
\end{equation}
while the $e$-foldings number is,
\begin{equation}\label{integraln3}
N=-\frac{6 \kappa ^2 \left(\frac{d \phi ^2}{2}-\frac{\kappa  \phi
^3}{3}\right)}{d \left(4 \kappa ^4 \lambda
+3\right)}\Big{|}_{\phi_f}^{\phi_i}\, .
\end{equation}
This model can be compatible with the ACT data as we now evince,
for various choices of the free parameters of the model. For
example the compatibility with the ACT data comes for the choice
$N=55$ and $(d,\lambda,M)=(1.2,10^{-18},2.21\times 10^{-10})$, in
which case we get $n_{\mathcal{S}}=0.9750$, $r= 0.006329$ and the
gravitational wave speed is $|c_T^2-1|=3.1487\times 10^{-17}$ and
also the amplitude of the scalar perturbations is
$\mathcal{P}_{\zeta}(k_*)=2.196\times 10^{-9}$.
\begin{figure}
\centering
\includegraphics[width=18pc]{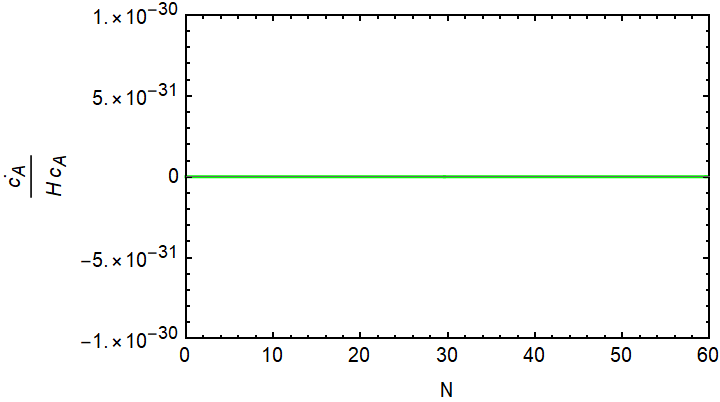}
\includegraphics[width=18pc]{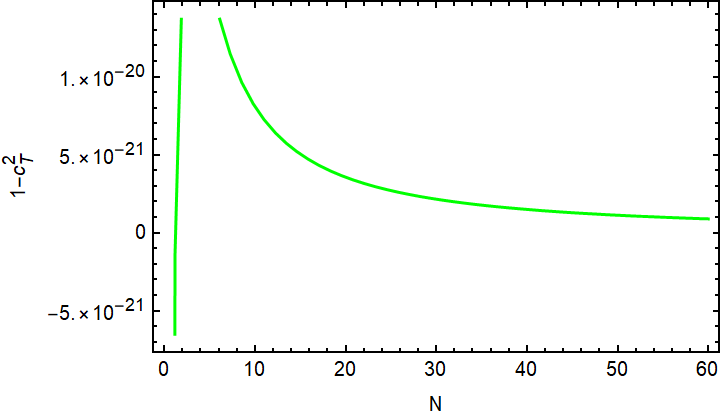}
\caption{The adiabaticity constraint $\frac{\dot{c}_A}{c_A H}$
(left plot) and the gravitational wave speed (right plot) as
functions of the $e$-foldings number for the model
(\ref{viablepotentials3}).}\label{plot4}
\end{figure}
Also the adiabaticity constraint is satisfied for the whole
duration of the inflationary era. This can be seen in Fig.
\ref{plot4} where we present the behavior of the constraint
$\frac{\dot{c}_A}{c_A H}$ as a function of the $e$-foldings number
(left plot) and the gravitational wave speed as a function of the
$e$-foldings (right plot). As it can be seen, the resulting theory
is phenomenologically viable and it also produces adiabatic
perturbations, for the whole inflationary era. The numerical
values of $\frac{\dot{c}_A}{c_A H}$ are too small, so we quote
here the actual values at the end and the beginning of inflation
and these are $\frac{\dot{c}_A}{c_A
H}\Big{|}_{\phi_f}=-8.39668\times 10^{-35}$ and
$\frac{\dot{c}_A}{c_A H}\Big{|}_{\phi_i}=-2.4693\times 10^{-44}$
respectively.

Now let us proceed with a model that is ACT only compatible, in
which case the scalar potential is chosen as follows,
\begin{equation}\label{viablepotentials4}
V(\phi)=M \left(1-\frac{\delta }{\kappa  \phi }\right)^{3/2}\, ,
\end{equation}
and $\xi'(\phi)$ in this case is,
\begin{equation}\label{xiphi4}
\xi'(\phi)=\frac{3 \delta  \lambda }{2 \kappa  M \phi ^2
\left(1-\frac{\delta }{\kappa  \phi }\right)^{5/2}}\, ,
\end{equation}
and the first slow-roll index is,
\begin{equation}\label{slowrollindexena4}
\epsilon_1=\frac{\delta ^2 \left(4 \kappa ^4 \lambda
+3\right)^2}{8 \kappa ^2 \phi ^2 (\delta -\kappa  \phi )^2}\, ,
\end{equation}
while the $e$-foldings number is,
\begin{equation}\label{integraln4}
N=-\frac{2 \kappa ^2 \left(\frac{\delta  \phi ^2}{2}-\frac{\kappa
\phi ^3}{3}\right)}{\delta  \left(4 \kappa ^4 \lambda
+3\right)}\Big{|}_{\phi_f}^{\phi_i}\, .
\end{equation}
This model can be compatible with the ACT data only as we now
evince, for various choices of the free parameters of the model.
For example the compatibility with the ACT data comes for the
choice $N=60$ and $(\delta,\lambda,M)=(1.7,10^{-18},6.35\times
10^{-10})$, in which case we get $n_{\mathcal{S}}=0.97626$, $r=
0.028628$ and the gravitational wave speed is
$|c_T^2-1|=1.3295\times 10^{-17}$ and also the amplitude of the
scalar perturbations is $\mathcal{P}_{\zeta}(k_*)=2.196\times
10^{-9}$.
\begin{figure}
\centering
\includegraphics[width=18pc]{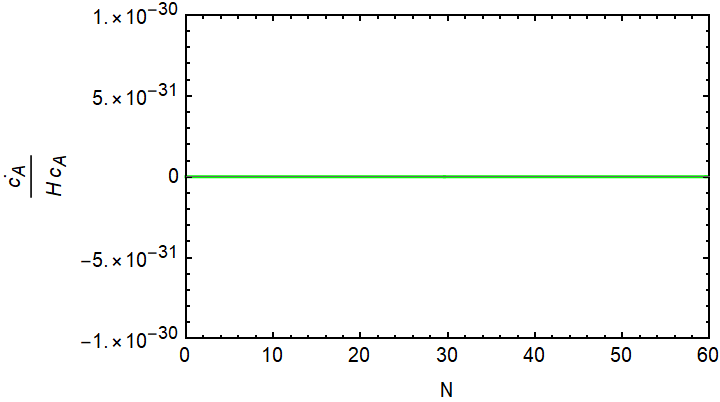}
\includegraphics[width=18pc]{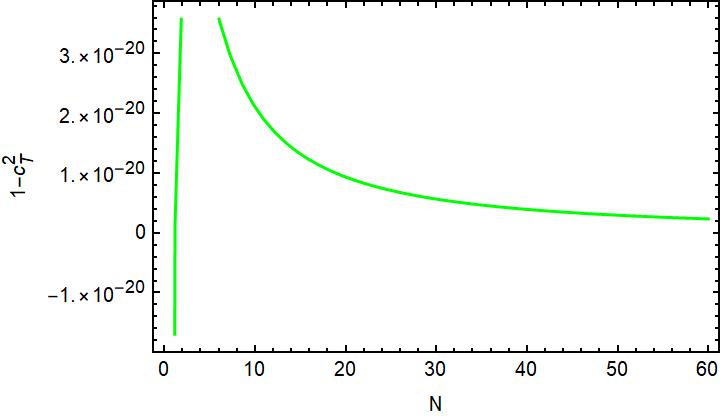}
\caption{The adiabaticity constraint $\frac{\dot{c}_A}{c_A H}$
(left plot) and the gravitational wave speed (right plot) as
functions of the $e$-foldings number for the model
(\ref{viablepotentials4}), taking into account the Planck data
compatibility of the model.}\label{plot5}
\end{figure}
Also the adiabaticity constraint is satisfied for the whole
duration of the inflationary era. This can be seen in Fig.
\ref{plot5} where we present the behavior of the constraint
$\frac{\dot{c}_A}{c_A H}$ as a function of the $e$-foldings number
(left plot) and the gravitational wave speed as a function of the
$e$-foldings (right plot). As it can be seen, the resulting theory
is phenomenologically viable and it also produces adiabatic
perturbations, for the whole inflationary era. The numerical
values of $\frac{\dot{c}_A}{c_A H}$ are too small, so we quote
here the actual values at the end and the beginning of inflation
and these are $\frac{\dot{c}_A}{c_A
H}\Big{|}_{\phi_f}=-4.0385\times 10^{-35}$ and
$\frac{\dot{c}_A}{c_A H}\Big{|}_{\phi_i}=-7.99588\times 10^{-44}$
respectively.

Finally we present another model that is both ACT and Planck
compatible, in which case the scalar potential is chosen as
follows,
\begin{equation}\label{viablepotentials5}
V(\phi)=\frac{M (\kappa  \phi )^2}{d+(\kappa  \phi )^2}\, ,
\end{equation}
and $\xi'(\phi)$ in this case is,
\begin{equation}\label{xiphi5}
\xi'(\phi)=\frac{2 d \lambda }{\kappa ^2 M \phi ^3}\, ,
\end{equation}
and the first slow-roll index is,
\begin{equation}\label{slowrollindexena5}
\epsilon_1=\frac{2 d^2 \left(4 \kappa ^4 \lambda +3\right)^2}{9
\kappa ^2 \phi ^2 \left(d+\kappa ^2 \phi ^2\right)^2}\, ,
\end{equation}
while the $e$-foldings number is,
\begin{equation}\label{integraln5}
N=\frac{\frac{3}{2} d \kappa ^2 \phi ^2+\frac{3 \kappa ^4 \phi
^4}{4}}{8 d \kappa ^4 \lambda +6 d}\Big{|}_{\phi_f}^{\phi_i}\, .
\end{equation}
This model can be compatible with both the ACT data and the Planck
data as we now evince, for various choices of the free parameters
of the model. For example the compatibility with the ACT data
comes for the choice $N=60$ and
$(d,\lambda,M)=(1.1,10^{-20},1.11\times 10^{-10})$, in which case
we get $n_{\mathcal{S}}=0.97432$, $r= 0.0066397$ and the
gravitational wave speed is $|c_T^2-1|=1.5891\times 10^{-19}$ and
also the amplitude of the scalar perturbations is
$\mathcal{P}_{\zeta}(k_*)=2.196\times 10^{-9}$. With regard to the
compatibility of the model with respect to the Planck data, the
compatibility with the Planck data comes for the choice $N=50$ and
$(d,\lambda,M)=(5.1,10^{-20},3.5\times 10^{-10})$, in which case
we get $n_{\mathcal{S}}=0.96809$, $r= 0.019747$ and the
gravitational wave speed is $|c_T^2-1|=9.3445\times 10^{-20}$ and
also the amplitude of the scalar perturbations is
$\mathcal{P}_{\zeta}(k_*)=2.196\times 10^{-9}$.
\begin{figure}
\centering
\includegraphics[width=18pc]{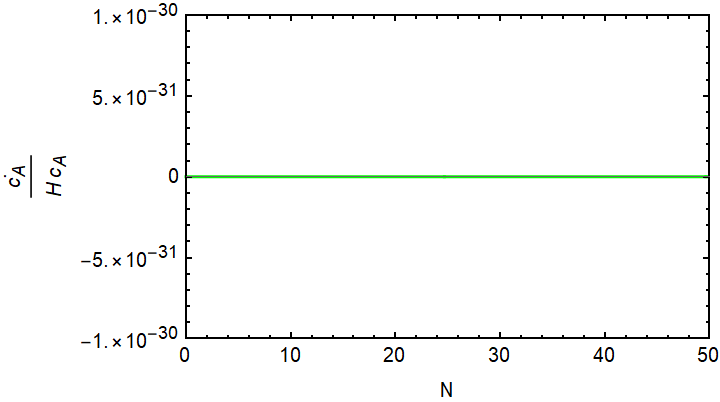}
\includegraphics[width=18pc]{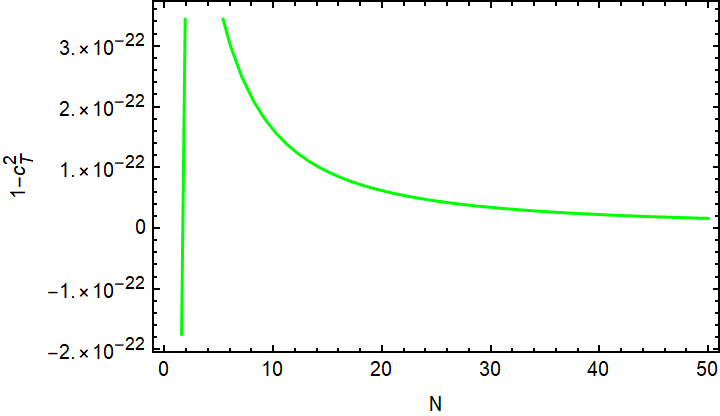}
\caption{The adiabaticity constraint $\frac{\dot{c}_A}{c_A H}$
(left plot) and the gravitational wave speed (right plot) as
functions of the $e$-foldings number for the model
(\ref{viablepotentials5}), taking into account the Planck data
compatibility of the model.}\label{plot6}
\end{figure}
Also the adiabaticity constraint is satisfied for the whole
duration of the inflationary era. This can be seen in Fig.
\ref{plot6} where we present the behavior of the constraint
$\frac{\dot{c}_A}{c_A H}$ as a function of the $e$-foldings number
(left plot) and the gravitational wave speed as a function of the
$e$-foldings (right plot). As it can be seen, the resulting theory
is phenomenologically viable and it also produces adiabatic
perturbations, for the whole inflationary era. The numerical
values of $\frac{\dot{c}_A}{c_A H}$ are too small, so we quote
here the actual values at the end and the beginning of inflation
and these are $\frac{\dot{c}_A}{c_A
H}\Big{|}_{\phi_f}=-2.49187\times 10^{-39}$ and
$\frac{\dot{c}_A}{c_A H}\Big{|}_{\phi_i}=-8.29926\times 10^{-48}$
respectively, by taking the Planck compatibility values of the
free parameters.

\section{Constrained EGB Theories, Non-Adiabaticity, Particle Production and New Features: Sinusoidal Power-Spectrum}

There is a second formalism to obtain a viable inflationary EGB
theory which is compatible with the GW170817 event, and it was
developed in Refs.
\cite{Oikonomou:2021kql,Oikonomou:2022xoq,Odintsov:2020sqy}. In
the context of this formalism, one utilizes the fact that the
gravitational wave speed of Eq. (\ref{GW}) is equal to unity when
$Q_f=0$, and this happens when $\ddot\xi=H\dot\xi$. This
constraint on the Gauss-Bonnet coupling can be expressed in terms
of the scalar field as follows,
\begin{equation}
\label{constraintupdatedmodel1} \centering
\xi''\dot\phi^2+\xi'\ddot\phi=H\xi'\dot\phi\, ,
\end{equation}
where the ``prime'' indicates differentiation with respect to the
scalar field. By assuming,
\begin{equation}\label{firstslowrollupdatedmodel}
 \xi'\ddot\phi \ll\xi''\dot\phi^2\, ,
\end{equation}
which is motivated by the scalar field slow-roll conditions, Eq.
(\ref{constraintupdatedmodel1}) takes the form,
\begin{equation}
\label{constraintupdatedmodel} \centering
\dot{\phi}\simeq\frac{H\xi'}{\xi''}\, .
\end{equation}
Eqs. (\ref{motion3}) and (\ref{constraintupdatedmodel}) yield,
\begin{equation}
\label{motion4updatedmodel} \centering
\frac{\xi'}{\xi''}\simeq-\frac{1}{3 H^2}\left(V'+12
\xi'H^4\right)\, .
\end{equation}
We focus on models that further satisfy the following,
\begin{equation}\label{mainnewassumptionupdatedmodel}
\kappa \frac{\xi '}{\xi''}\ll 1\, ,
\end{equation}
and in addition,
\begin{equation}\label{scalarfieldslowrollextraupdatedmodel}
12 \dot\xi H^3=12 \frac{\xi'^2H^4}{\xi''}\ll V\, ,
\end{equation}
which is strongly related to the constraint
(\ref{mainnewassumptionupdatedmodel}). Eqs.
(\ref{constraintupdatedmodel}) and
(\ref{scalarfieldslowrollextraupdatedmodel}), yield,
\begin{equation}
\label{motion5updatedmodel} \centering
H^2\simeq\frac{\kappa^2V}{3}\, ,
\end{equation}
\begin{equation}
\label{motion6updatedmodel} \centering \dot
H\simeq-\frac{1}{2}\kappa^2 \dot\phi^2\, ,
\end{equation}
\begin{equation}
\label{motion8updatedmodel} \centering
\dot\phi\simeq\frac{H\xi'}{\xi''}\, .
\end{equation}
Also Eqs. (\ref{motion5updatedmodel}) and
(\ref{scalarfieldslowrollextraupdatedmodel}) yield,
\begin{equation}\label{mainconstraint2updatedmodel}
 \frac{4\kappa^4\xi'^2V}{3\xi''}\ll 1\, .
\end{equation}
Thus the scalar field potential and the Gauss-Bonnet coupling are
not free to choose in this context, but these satisfy,
\begin{equation}
\label{maindiffeqnnewupdatedmodel} \centering
\frac{V'}{V^2}+\frac{4\kappa^4}{3}\xi'\simeq 0\, .
\end{equation}
So the slow-roll indices in this class of models take the form,
\begin{equation}
\label{index1updatedmodel} \centering
\epsilon_1\simeq\frac{\kappa^2
}{2}\left(\frac{\xi'}{\xi''}\right)^2\, ,
\end{equation}
\begin{equation}
\label{index2updatedmodel} \centering
\epsilon_2\simeq1-\epsilon_1-\frac{\xi'\xi'''}{\xi''^2}\, ,
\end{equation}
\begin{equation}
\label{index3updatedmodel} \centering \epsilon_3=0\, ,
\end{equation}
\begin{equation}
\label{index4updatedmodel} \centering
\epsilon_4\simeq\frac{\xi'}{2\xi''}\frac{\mathcal{E}'}{\mathcal{E}}\,
,
\end{equation}
\begin{equation}
\label{index5updatedmodel} \centering
\epsilon_5\simeq-\frac{\epsilon_1}{\lambda}\, ,
\end{equation}
\begin{equation}
\label{index6updatedmodel} \centering \epsilon_6\simeq
\epsilon_5(1-\epsilon_1)\, ,
\end{equation}
where, $\mathcal{E}=\mathcal{E}(\phi)$ and $\lambda=\lambda(\phi)$
is,
\begin{equation}\label{functionE}
\mathcal{E}(\phi)=\frac{1}{\kappa^2}\left(
1+72\frac{\epsilon_1^2}{\lambda^2} \right),\,\, \,
\lambda(\phi)=\frac{3}{4\xi''\kappa^2 V}\, .
\end{equation}
Hence the observational indices of inflation in this scenario are,
\begin{equation}
\label{spectralindex} \centering
n_{\mathcal{S}}=1-4\epsilon_1-2\epsilon_2-2\epsilon_4\, ,
\end{equation}
regarding the scalar spectral index, while the tensor spectral
index and the tensor-to-scalar ratio acquire very simple forms,
\begin{equation}\label{tensorspectralindexfinalupdatedmodel}
n_{\mathcal{T}}\simeq -2\epsilon_1\left ( 1-\frac{1}{\lambda
}+\frac{\epsilon_1}{\lambda}\right)\, ,
\end{equation}
\begin{equation}\label{tensortoscalarratiofinalupdatedmodel}
r\simeq 16\epsilon_1\, .
\end{equation}
Also the $e$-foldings number takes the form,
\begin{equation}
\label{efoldsupdatedmodel} \centering
N=\int_{t_i}^{t_f}{Hdt}=\int_{\phi_i}^{\phi_f}\frac{H}{\dot{\phi}}d\phi=\int_{\phi_i}^{\phi_f}{\frac{\xi''}{\xi'}d\phi}\,
.
\end{equation}
The phenomenology of such framework was thoroughly analyzed in
Refs. \cite{Oikonomou:2021kql,Oikonomou:2022xoq,Odintsov:2020sqy},
so let us consider a viable model that is both ACT and Planck
compatible, also considering the adiabaticity constraints.

Let us consider the model with scalar Gauss-Bonnet coupling
function,
\begin{equation}
\label{modelA} \xi(\phi)=\beta  (\kappa  \phi )^{\nu }\, ,
\end{equation}
where $\beta$ is some dimensionless constant and $\kappa=1/M_p$.
Combining (\ref{modelA}) and Eq.
(\ref{maindiffeqnnewupdatedmodel}) we obtain,
\begin{equation}
\label{potA} \centering V(\phi)=\frac{3}{4 \beta  \kappa ^{\nu +4}
\phi ^{\nu }+3 \gamma  \kappa ^4} \, ,
\end{equation}
where $\gamma$ is some dimensionless integration constant. Then we
get,
\begin{equation}
\label{index1A} \centering \epsilon_1\simeq \frac{\kappa ^2 \phi
^2}{2 (\nu -1)^2} \, ,
\end{equation}
\begin{equation}
\label{index2A} \centering \epsilon_2\simeq -\frac{\kappa ^2 \phi
^2-2 \nu +2}{2 (\nu -1)^2}\, ,
\end{equation}
\begin{equation}
\label{index3A} \centering \epsilon_3=0\, ,
\end{equation}
\begin{equation}
\label{index4A} \centering \epsilon_4\simeq \frac{\phi
\left(\kappa \, (2 \nu -4) \alpha (\phi ) \zeta (\phi )-8 \beta
\nu \zeta (\phi ) \kappa ^{\nu +5} \phi ^{\nu }\right)}{2 \kappa
(\nu -1) \phi  \alpha (\phi ) (\zeta (\phi )+1)} \, ,
\end{equation}
\begin{equation}
\label{index5A} \centering \epsilon_5\simeq -\frac{2 \beta \,
\kappa^4\, \nu (\kappa  \phi )^{\nu }}{(\nu -1)  \alpha (\phi )}
\, ,
\end{equation}
\begin{equation}
\label{index6A} \centering \epsilon_6\simeq -\frac{\beta \,
\kappa^4\, \nu (\kappa  \phi )^{\nu } \left(-\kappa ^2 \phi ^2+2
\nu ^2-4 \nu +2\right)}{(\nu -1)^3 \alpha (\phi )} \, .
\end{equation}
So upon solving $\epsilon_1\simeq \mathcal{O}(1)$ we get
$\phi_f\simeq \frac{\sqrt{2} (\nu-1)}{\kappa }$, and by solving
Eq. (\ref{efoldsupdatedmodel}) with respect to the variable
$\phi_i$ we obtain $\phi_i=\frac{\sqrt{2} (\nu-1) e^{-\frac{N}{\nu
-1}}}{\kappa }$. Hence the scalar spectral index takes the form,
\begin{align}\label{spectralpowerlawmodelupdatedmodel}
n_{\mathcal{S}}\simeq -1+\frac{2 (\nu -2)}{\nu -1}-2\, e^{-\frac{2
N}{\nu -1}} + (\nu -1)^{2 \nu -3} \nu ^2  \frac{9\, \beta ^2
\,2^{\nu +6} \,\left(\beta \, 2^{\frac{\nu }{2}+3}\, (\nu -1)^{\nu
+1}+3 \gamma  (\nu -2) e^{\frac{\nu N}{\nu
-1}}\right)}{\left(\beta \, 2^{\frac{\nu }{2}+2}\, (\nu -1)^{\nu
}+3 \gamma  e^{\frac{\nu  N}{\nu -1}}\right)^3} \, ,
\end{align}
and also the tensor-to-scalar ratio becomes,
\begin{equation}\label{tensortoscalarfinalmodelpowerlawupdatedmodel}
r\simeq 16\, e^{-\frac{2 N}{\nu -1}}\, .
\end{equation}
In addition, the tensor-spectral index takes the form,
\begin{align}\label{tensorspectralindexpowerlawmodelupdatedmodel}
n_{\mathcal{T}}\simeq \frac{12 \gamma  e^{\frac{(\nu -4) N}{\nu
-1}}}{\beta \, 2^{\frac{\nu }{2}+2} (\nu -1)^{\nu }+3 \gamma
e^{\frac{\nu  N}{\nu -1}}} -\frac{\beta \, 2^{\frac{\nu }{2}+3}
(\nu -1)^{\nu -1} e^{-\frac{4 N}{\nu -1}} \left(-3 \nu +(\nu -1)
\nu  e^{\frac{4 N}{\nu -1}}+2\right)}{\beta\, 2^{\frac{\nu }{2}+2}
(\nu -1)^{\nu }+3 \gamma  e^{\frac{\nu N}{\nu -1}}} \, .
\end{align}
A viable set of parameter values that yields compatibility with
the ACT data is ($\beta, \gamma,\nu)=(2.89912\times
10^6,10.6\times 10^8,19.98)$ for $N=60$, for which
$n_\mathcal{S}=0.973503$, $n_{\mathcal{T}}=0.0442981$,
$r=0.0287288$ and all these observables are compatible with the
ACT constraints and the updated Planck tensor-to-scalar ratio
constraints. For these values, the amplitude of the scalar
perturbations is $\mathcal{P}_{\zeta}(k)=2.19\times 10^{-9}$ which
is of course by construction, compatible with the Planck 2018
constraints \cite{Planck:2018jri}. Regarding the Planck data
constraints, by choosing $(\nu,\gamma,\beta)=(20,5768,-13.191)$,
we get, $n_{\mathcal{S}}=0.966$ and $r=0.0289206$ and
$\mathcal{P}_{\zeta}(k_*)=2.19673\times 10^{-9}$.
\begin{figure}
\centering
\includegraphics[width=18pc]{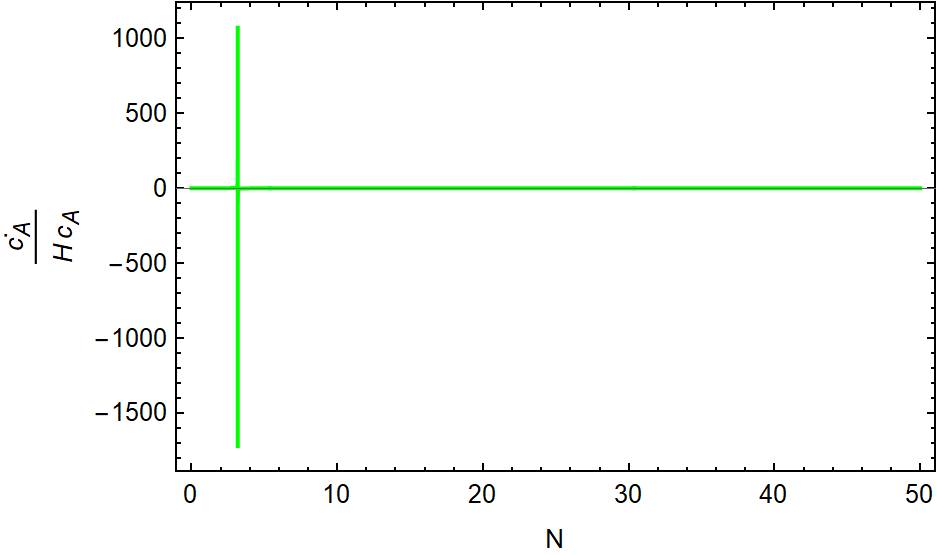}
\includegraphics[width=18pc]{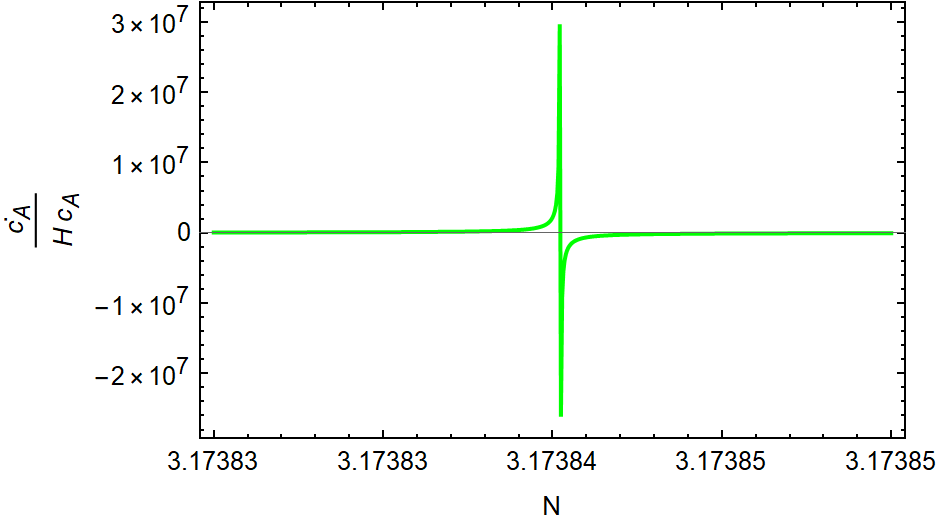}
\includegraphics[width=18pc]{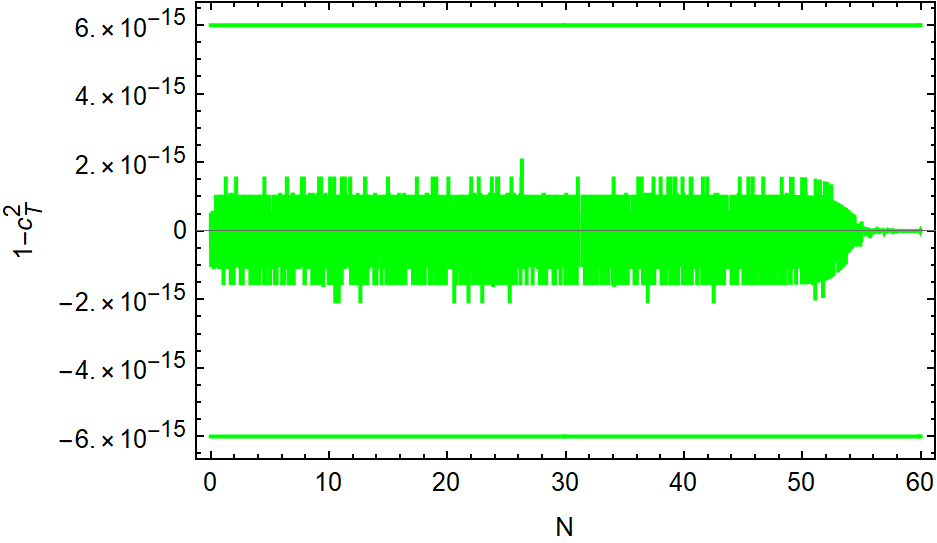}
\caption{The adiabaticity constraint $\frac{\dot{c}_A}{c_A H}$ for
the whole duration of inflation (upper left plot), and for the
$e$-foldings range $N=[3.17383, 3.17385]$ (upper right plot), and
the gravitational wave speed (bottom plot) as functions of the
$e$-foldings number for the model (\ref{modelA}), taking into
account the Planck data compatibility of the model.}\label{plot7}
\end{figure}
Also the adiabaticity constraint is satisfied for the largest part
of the inflationary era, but it is strongly violated in the last
few $e$-foldings. Indeed, some characteristic values are
$\frac{\dot{c}_A}{c_A H}\Big{|}_{\phi_f}=-3.72897\times 10^{-16}$
at the end of inflation,  $\frac{\dot{c}_A}{c_A
H}\Big{|}_{\phi_i}=-0.000676488$ at the beginning of inflation and
$\frac{\dot{c}_A}{c_A H}=-19.0845$ for $N=3.1$. Hence we need a
closer look on the behavior of the adiabaticity condition
$\frac{\dot{c}_A}{c_A H}$, so in Fig. \ref{plot7} we present the
behavior of the constraint $\frac{\dot{c}_A}{c_A H}$ as a function
of the $e$-foldings number in both left and right upper plots. The
right plot is a close up of the left plot, for small $e$-foldings
number. As it can be seen, the theory violates the adiabaticity
constraints in the last few $e$-foldings. In the bottom plot of
Fig. \ref{plot7} we present the behavior of the gravitational wave
speed as a function of the $e$-foldings, including the upper and
lower limits of the GW170817 constraints. As it can be seen, the
gravitational wave constraints are well respected.

Since the theory becomes non-adiabatic, interesting phenomena are
expected, so let us first analyze possible reasons why this
unexpected non-adiabaticity occurs for this theoretical framework.
Using the analytic form of the sound wave speed $c_A$ for EGB in
Eq. (\ref{soundspeed}), we end up to the following expression,
\begin{equation}\label{expressionfinaloized}
\frac{\dot{c}_A}{c_A H}\sim \mathcal{O}\left(
\frac{\kappa^2\left(\ddot{\xi}H+\dot{\xi}\dot{H}\right)}
{H^2\epsilon_1} \right)\, .
\end{equation}
Using the definition of the first slow-roll index,
\begin{equation}
\epsilon_1=-\frac{\dot{H}}{H^2}\, ,
\end{equation}
the above expression (\ref{expressionfinaloized}) can be rewritten
as follows,
\begin{equation}\label{asxeton}
\frac{\dot{c}_A}{c_A H}\sim \mathcal{O}\left(
\frac{\kappa^2H\dot{\xi}}{\epsilon_1} \left(
\frac{\ddot{\xi}}{H\dot{\xi}}-\epsilon_1 \right) \right)\, ,
\end{equation}
therefore giving a rough estimate, the condition for adiabaticity
of the scalar perturbations in EGB is,
\begin{equation}\label{finalconditionsadiabaticitycriteria}
\left|\frac{\ddot{\xi}}{\dot{\xi}H}\frac{\kappa^2\dot{\xi}H}{\epsilon_1}\right|\ll
1\, .
\end{equation}
Notice that $\dot{\xi}H$ enters in the field equation and thus the
condition $\left|\kappa^2\dot{\xi}H\right|\ll 1$ is an effective
field theory requirement. Before proceeding, let us here reveal an
important feature regarding the adiabaticity criteria. The
requirement $\left|\frac{\ddot{\xi}}{\dot{\xi}H}\right|\ll 1$,
contributes to the adiabaticity criterion
(\ref{finalconditionsadiabaticitycriteria}), and can be written
as,
\begin{equation}\label{eq1}
\frac{\ddot{\xi}}{\dot{\xi}H}=\frac{\xi''\dot{\phi}}{\xi'H}+\frac{\ddot{\phi}}{H\dot{\phi}}\,
,
\end{equation}
where the ``prime'' denotes here differentiation with respect to
the scalar field. In view of Eq. (\ref{eq1}), the adiabaticity
requirement reads,
\begin{equation}\label{eq2}
\left|\frac{\xi''\dot{\phi}}{\xi'H}+\frac{\ddot{\phi}}{H\dot{\phi}}\right|\ll
1\, ,
\end{equation}
therefore, in order to have adiabatic perturbations, one must have
simultaneously,
\begin{equation}\label{siumadiabat}
\kappa^2\frac{\xi''}{\xi'}\ll 1,\,\,\,\frac{\dot{\phi}}{H}\ll
1,\,\,\,\frac{\ddot{\phi}}{H\dot{\phi}}\ll 1\, .
\end{equation}
The condition $\frac{\dot{\phi}}{H}\ll 1$ can be automatically
satisfied if $\dot{\phi}=\gamma H^{-m}$ with $m>0$, and such
tracking conditions have been used in the literature
\cite{Oikonomou:2026mvp} and it produces analytic solutions for
inflation in single scalar field theory. A simple extension of the
tracking condition in the context of EGB can therefore produce
automatically adiabatic perturbations, if
$\kappa^2\frac{\xi''}{\xi'}\ll 1$ and
$\frac{\ddot{\phi}}{H\dot{\phi}}\ll 1$, with the later being
simply a slow-roll condition on the scalar field.

Now in the present context we have $\ddot{\xi}\sim H\dot{\xi}$ by
construction, so the adiabaticity condition in Eq.
(\ref{siumadiabat}) might be violated due to this and the relative
fraction $\sim \frac{\dot{\xi}H}{\epsilon_1}$. This is probably
the source of the non-adiabaticity of this framework, but note
that the conditions (\ref{siumadiabat}) are not sufficient to
determine the adiabaticity of the theory. Only the numerical study
appearing in Fig. \ref{plot7} provides enough and solid details on
the problem of adiabaticity.

Now, since the theory becomes non-adiabatic at the end of
inflation, interesting phenomena may occur at the end of
inflation. The adiabaticity conditions are violated at the end of
the inflationary regime and in fact in the very last few
$e$-foldings, so at the end of the inflationary regime, the
Bogoliubov coefficient $\beta_k$ in Eq. (\ref{betakcosmictime})
will increase, leading to particle production at the end of the
inflationary regime. The energy density of these particles will
be,
\begin{equation}\label{energythermal}
\rho_{pp}=\frac{1}{2\pi^2a^4}\int \mathrm{d}k
k^3\omega_k(t)|\beta_k|^2\, ,
\end{equation}
so by considering only the subhorizon modes that nearly exit the
horizon at the end of inflation, we have $\omega_k\sim c_A k$, and
since $c_A\sim 1$ for all the models we analyzed, the final energy
density of the produced particles is,
\begin{equation}\label{finalenergydensiy}
\rho_{pp}=\frac{1}{2\pi^2a^4}\int \mathrm{d}k k^4|\beta_k|^2\, .
\end{equation}
These particles redshift as radiation, hence this particle
production at the end of the inflationary era is a mechanism for
preheating which will facilitate the reheating era. Hence, this
class of EGB inflationary models may lead to particle production
at the end of inflation which acts as a gravitational reheating
mechanism. Note that for the non-adiabatic modes, the CMB is not
affected, since the non-adiabatic modes exit the Hubble horizon at
the end of inflation, nearly at $N\simeq 3$. However, we may have
some secondary gravitational wave effects, which we aim to further
examine in a future work.

Let us reveal another interesting feature of this class of models,
suppose that in the context of this section's scenario, the
inflationary regime commenced from a Bunch-Davies vacuum and the
``out'' state is given by Eq. (\ref{outstate}). The power spectrum
in this scenario is,
\begin{equation}\label{powerspectrumgeneral}
P(k)=
\frac{k^3}{2\pi^2}\frac{1}{z^2}|u_k|^2=\frac{k^3}{2\pi^2}\frac{1}{z^2}\frac{1}{2k}\left(|a_k|^2+|\beta_k|^2+a_k\beta_k^*e^{-2ik\eta}+a_k^*\beta_k\,e^{2ik\eta}\,
, \right)
\end{equation}
which can be written as follows,
\begin{equation}\label{powerspectrumgeneral1}
P(k)=\frac{k^3}{2\pi^2}\frac{1}{z^2}\frac{1}{2k}\left(2
\mathrm{Re}\left(a_k\beta_k^*e^{-2ik\eta}
\right)+|a_k|^2+|\beta_k|^2\right)\, ,
\end{equation}
or equivalently,
\begin{equation}\label{powerspec3}
P(k)=\frac{k^3}{2\pi^2}\frac{1}{z^2}\frac{1}{2k}\left(2|a_k||\beta_k|\cos\left(
2k\eta+\phi_k\right)+|a_k|^2+|\beta_k|^2\right)\, ,
\end{equation}
where $\phi_k=\mathrm{arg}(a_k)-\mathrm{arg}(\beta_k)$. The
Bogoliubov coefficients $a_k$ and $\beta_k$ satisfy,
\begin{equation}\label{bogoliubovnorm}
|a_k|^2-|\beta_k|^2=1\, ,
\end{equation}
so the power spectrum becomes,
\begin{equation}\label{finalpowerspectrum0}
P(k)=P_0(k)\left(1+2|\beta_k|^2+2\sqrt{1+|\beta_k|^2}|\beta_k|\cos\left(
2k\eta+\phi_k\right)\right)\, ,
\end{equation}
where we used the unperturbed power spectrum of the Bunch-Davies
vacuum $P_0(k)=\frac{k^3}{2\pi^2}\frac{1}{z^2}\frac{1}{2k}$. Hence
the power spectrum of the scalar perturbations in this case is
rendered an oscillating spectrum, which is also known to appear in
the literature for exotic inflationary scenarios, see for example
\cite{Zeng:2018ufm,Domenech:2019cyh,Danielsson:2002kx,Jackson:2013vka,Kempf:2000ac,Easther:2001fz,Martin:2003kp,Chen:2008wn,Flauger:2009ab,Covi:2006ci,Hamann:2007pa,
Hazra:2010ve,Shafieloo:2003gf,Achucarro:2013cva,Hazra:2014jwa,Chen:2014cwa,Nicholson:2009pi,Hunt:2015iua,Braglia:2021ckn,Braglia:2021sun,Antony:2021bgp,
Hazra:2022rdl,Antony:2022ert,Silverstein:2008sg,McAllister:2008hb,Behbahani:2011it,Wang:2002hf,Easther:2002xe,Bozza:2003pr}.
This is a new feature for the EGB inflationary theories, but note
that this oscillating feature is acquired by the power spectrum
only at the last few $e$-foldings, so it surely does not affect
the CMB modes, and is surely distinct from the scenarios
\cite{Zeng:2018ufm,Domenech:2019cyh,Danielsson:2002kx,Jackson:2013vka,Kempf:2000ac,Easther:2001fz,Martin:2003kp,Chen:2008wn,Flauger:2009ab,Covi:2006ci,Hamann:2007pa,
Hazra:2010ve,Shafieloo:2003gf,Achucarro:2013cva,Hazra:2014jwa,Chen:2014cwa,Nicholson:2009pi,Hunt:2015iua,Braglia:2021ckn,Braglia:2021sun,Antony:2021bgp,
Hazra:2022rdl,Antony:2022ert,Silverstein:2008sg,McAllister:2008hb,Behbahani:2011it,Wang:2002hf,Easther:2002xe,Bozza:2003pr},
because in these references the oscillating feature occurs for the
whole inflationary era and not for a few $e$-foldings in the end
of inflation. Now it would be interesting to check for secondary
gravitational wave effects at the end of inflation, due to this
oscillating behavior, and we aim to investigate this feature in a
future work.

\section{Conclusions}

In this work we considered the adiabaticity of the cosmological
scalar and tensor perturbations of EGB theories of gravity. We
discussed the quantitative effects of the adiabaticity conditions
and how do these affect the quantities entering in the EGB
Lagrangian. We considered two classes of inflationary EGB theories
of gravity which are compatible with the current CMB experiments
and are also compatible with the GW170817 event. In the first
class, the resulting cosmological perturbations are adiabatic
during inflation, save the few last $e$-foldings, and this results
to particle creation at the end of inflation and also the power
spectrum is affected acquiring sinusoidal terms. Also this
adiabaticity violation at the end of inflation generates particle
creation and thus the preheating is facilitated in this model. The
other class of EGB theories has adiabatic perturbations during the
inflationary regime.  An interesting observation regarding the
adiabaticity criterion of EGB theories is that the adiabaticity of
EGB theory can occur solely if $\dot{\phi}=f(H)$. Such tracking
conditions have been analyzed in the literature
\cite{Oikonomou:2026mvp}, so it would be intriguing to work such a
scenario in some detail. We aim to address this topic in a future
work.

In retrospect, from the two viable EGB theories we presented, the
unconstrained EGB theories are more stable and free from
adiabaticity pathologies. For these theories, the dark energy
perspective would be quite interesting to address, since in the
present context, in the unconstrained EGB theories, the scalar
Gauss-Bonnet coupling function $\xi(\phi)$ satisfies
$\xi(\phi)=-\frac{\lambda}{V(\phi)}$. Thus inflation in the this
context starts when $\kappa^2 V\gg 1$ and when $\xi(\phi)\ll 1$,
thus as inflation ends, the Gauss-Bonnet coupling increases, hence
the dark energy era will be affected by the increasing
Gauss-Bonnet coupling function $\xi(\phi)$. We aim to work along
this research line in the near future.

\end{document}